\documentclass[useAMS,usenatbib]{mn2e}

\pdfoutput=1

\usepackage{mathrsfs}
\usepackage{txfonts}
\usepackage[T1]{fontenc}
\usepackage{aecompl}
\usepackage{graphicx}
\usepackage{array}

\usepackage{amsmath,amssymb}
\usepackage[english]{babel}
\usepackage{enumerate}
\usepackage{xspace}

\usepackage[usenames,dvipsnames]{color}

\def\ie{{\rm i.e.}\xspace}
\def\eg{{\rm e.g.}\xspace}

\def\cf{{\rm cf.}\xspace}
\def\be{\begin{equation}}
\def\ee{\end{equation}}
\def\bea{\begin{eqnarray}}
\def\eea{\end{eqnarray}}
\def\q#1{\lq{#1}\rq}
\newcommand{\sect}[1]{Sect.~\ref{#1}}

\newcommand{\fig}[1]{Fig.~\ref{#1}}

\newcommand{\eq}[1]{Eq.~(\ref{#1})}

\newcommand{\tab}[1]{Table~\ref{#1}}
\newcommand{\citeg}[1]{\citep[\eg][]{#1}}
\newcommand{\refeq}[1]{Eq.~(\ref{eq:#1})\xspace}

\newcommand{\nn}{\nonumber}
\newcommand{\cl}{\ensuremath{C_\ell}}
\newcommand{\alm}{\ensuremath{a_{lm}}}
\newcommand{\clth}{\ensuremath{\cl^\textrm{th}}} 
\newcommand{\fsky}{\ensuremath{f_\textrm{sky}}\xspace}
\newcommand{\fskyc}{\ensuremath{\fsky^{A \times B}} }
\newcommand{\lik}{\ensuremath{\mathscr{L}}}
\newcommand{\lmin}{\ensuremath{\ell_\textrm{min}}}
\newcommand{\lmax}{\ensuremath{\ell_\textrm{max}}}
\newcommand{\sa}{\sigma_A}
\renewcommand{\sb}{\sigma_B}
\newcommand{\Na}{\ensuremath{N^A_\ell}\xspace}
\newcommand{\Nb}{\ensuremath{N^B_\ell}\xspace}
\newcommand{\Nc}{\ensuremath{N^C_\ell}\xspace}
\newcommand{\clAB}{\ensuremath{\hat \cl^{A \times B}}\xspace}
\newcommand{\pdf}{p.d.f\xspace}
\providecommand{\abs}[1]{\lvert#1\rvert}
\newcommand{\E}[1]{\ensuremath{E\left[#1\right]}}
\newcommand\planck{\emph{Planck}}
\newcommand\wmap{\emph{WMAP}}
\newcommand\hfi{\emph{Planck}-HFI}
\newcommand\lfi{\emph{Planck}-70} 

\newcommand\hfic{\emph{Planck}-100} 
\newcommand\hficq{\emph{Planck}-143} 
\newcommand{\colorcode}{\wmap$\times$\lfi = orange, \wmap$\times$\planck-100\ = gold, \wmap$\times$\planck-143\ = purple, \lfi$\times$\planck-100\ = red, \lfi$\times$\planck-143\ = blue, \planck-100$\times$\planck-143\ = green}

\newcommand\tauvp{0.078}

\def\apss{\ref@jnl{Ap\&SS}}             
\def\aapr{\ref@jnl{A\&A~Rev.}}          
\def\aaps{\ref@jnl{A\&AS}}              
\def\pra{\ref@jnl{Phys.~Rev.~A}}        
\def\prb{\ref@jnl{Phys.~Rev.~B}}        
\def\prc{\ref@jnl{Phys.~Rev.~C}}        
\def\pre{\ref@jnl{Phys.~Rev.~E}}        
\def\prl{\ref@jnl{Phys.~Rev.~Lett.}}    
\def\plb{\ref@jnl{Phys.~Lett.~B}}       

\title[Large-scale CMB cross-spectra likelihoods]{Large-scale CMB temperature and polarization cross-spectra likelihoods}

\author[Mangilli, Plaszczynski, Tristram]{A. Mangilli$^{1,2}$, S. Plaszczynski$^{1}$ and M. Tristram$^{1}$\\
$^1${Laboratoire de l'Acc\'{e}l\'{e}rateur Lin\'{e}aire, Universit\'{e} Paris-Sud 11, CNRS/IN2P3, Orsay, France}\\
$^2${Institut d'Astrophysique Spatiale, B\^{a}t. 121, Universit\'{e} Paris-Sud 11, CNRS, Orsay, France}\\
}

\date{\today}

\begin{document}

\maketitle


\begin{abstract}
We present a cross-spectra based approach for the analysis of CMB data at large angular scales to constrain the reionization optical depth $\tau$, the tensor to scalar ratio $r$ and the amplitude of the primordial scalar perturbations $A_s$. 
With respect to the pixel-based approach developed so far, using cross-spectra has the unique advantage to eliminate spurious noise bias and to give a better handle over residual systematics, allowing to efficiently combine the cosmological information encoded in cross-frequency or cross-dataset spectra. We present two solutions to deal with the non-Gaussianity of the $\hat{C}_\ell$ estimator distributions at large angular scales: the first one relies on an analytical parametrization of the estimator distribution, 
while the second one is based on modification of the Hamimache\&Lewis likelihood approximation at large  angular scales.
The modified HL method (oHL) is powerful and complete. It allows to deal with multipole and mode correlations for a combined temperature and polarization analysis. 
We validate our likelihoods on numerous simulations that include the realistic noise levels of the \wmap, \planck-LFI and \planck-HFI experiments, demonstrating their validity over a broad range of cross-spectra configurations.
\end{abstract}



\section{Introduction}\label{Sec:intro}

One of the main challenges left for the present and future Cosmic
Microwave Background (CMB) experiments is the high precision
measurement of the CMB polarization anisotropies at large angular
scales.
This signal is extremely interesting because it encodes unique informations about the ionization history of the Universe and the inflationary epoch and it can be used as an independent and complementary probe to the small scale CMB information to constrain two important cosmological parameters: the optical depth to reionization $\tau$ and the tensor-to-scalar ratio parameter $r$ which is related to the amplitude of the primordial tensor modes. Moreover, the large scales CMB signal is useful in breaking parameter degeneracies, in particular concerning the two parameters: $\tau$ and  the amplitude of the primordial scalar density perturbations $A_s$ which are strongly correlated through the amplitude of the first acoustic peak $\mathcal{A}_{TT}=A_s e^{-2 \tau}$. %

Current CMB experiments, in particular the ones that, as \planck\  \citep{2015arXiv150201582P}, targeted the accurate measurement of the CMB temperature anisotropies, have now reached a level of precision and resolution such that they have exploited all their statistical power, and are now limited by the systematic effects related to the instrument design and technology. An unprecedented accuracy and care at each step of
the data analysis and its interpretation is therefore required to access the cosmological information encoded in the CMB polarization anisotropies at large angular scales. 
In this paper we address this issue focusing on the importance of
developing statistical methods specific to the analysis of CMB
data at large angular scales that allow to minimize the impact of residual systematics related to
the experimental configuration and design.

Given that the distribution of the CMB anisotropies is compatible with
a Gaussian distribution, all the relevant statistical information 
is encoded in the two points correlation function of the CMB
temperature and polarization anisotropies or, equivalently, its
projection in harmonic space: the angular power spectrum of the CMB
temperature and polarization fields.
This is defined as $\hat{C}_\ell=\langle a_{\ell m} a^*_{\ell' m'}\rangle \delta_{\ell \ell'}$, where $a_{\ell m}$ are the coefficients of the spherical harmonic decomposition. 
The connection between the measured CMB data and the theory is done
through the CMB likelihood function ${\lik}=P({\bf d}|C_\ell(\alpha))$
that quantifies the match between the CMB data {\bf d} and a given
theoretical model parametrized e.g. by a theoretical power spectrum
$C_\ell({\bf \alpha})$ defined in terms of a set of cosmological
parameters ${\bf \alpha}$.

 So far the analysis of the CMB anisotropies at large angular scales
has mostly been based on methods that relies on low resolution maps 
in order to compute the exact CMB likelihood function in pixel space,
${\lik}=P({\bf d} |C_\ell(\alpha))$, with ${\bf d}\equiv M({\bf
p})=\sum_{\ell m} a_{\ell m} Y_{\ell m}( {\bf p})$. 
This approach is based on the fact that, given that the CMB
anisotropies are compatible with a gaussian distribution with random
phases, the $a_{\ell m}$ follow a multi-variate Gaussian
distribution. The likelihood function, written in pixel space or,
equivantely, in terms of the $a_{\ell m}$ coefficients, is gaussian
and therefore can be computed exactly
\citep{Gorski:1994,Slosar:2004fr,Page:2006hz,WMAP9}.

The problem of this approach is that, in the case of a real CMB
experiment, the maps consist in a combination of signal, noise,
instrumental systematics and must account for the incomplete sky
coverage necessary to minimize the impact of the foregrounds
contamination. In order to achieve the required accuracy at large
angular scales, the noise matrix in pixel space must be reconstructed
with extremely high accuracy to avoid spurious bias on the parameters
reconstruction. However this accuracy can be extremely hard to achieve given
the difficulty of the precise characterization not only of the noise
but also of the residuals systematics related e.g. to the instrument,
the scanning strategy and the residual foregrounds.

Alternatively, the likelihood function could be defined in the
harmonic space as done e.g. in the small scales analysis where the
data compression from CMB maps to angular power spectra is necessary
for computational and numerical reasons. 
However, the complication of working in harmonic space at large
angular scales (low-$\ell$ multipoles) is 
related to the fact that the distribution of the $\hat{C}_\ell$ estimators at low-$\ell$ is non-Gaussian.
In harmonic space the $\hat{C}_\ell$ consist in the sum of the square of the harmonic coefficients $a_{\ell m}$ and they have a reduced-$\chi^2$ distribution.
Therefore the likelihood of a theoretical power spectrum as a function
of the measured $\hat{C}_\ell$ is non-Gaussian. Contrary to the
small-scales analysis, the CMB low-$\ell$ analysis is particularly
concerned by this issue given that the central limit theorem cannot be
invoked.
Previous studies, \citep{Percival:2006ss,Hamimeche:2008ai}, developed a CMB
analysis on large angular scales based on the likelihood definition in
harmonic space in terms of auto-spectra, that is to say CMB angular
power spectra obtained from a given single frequency/dataset CMB
map. This approach however shared problems similar to the pixel based
likelihood approach, in particular in terms of the dependency to the noise and of
the accurate characterization of the systematics effects at the
auto-spectra levels.

In this paper we propose to extend the cross-spectra based approach
for the analysis of the CMB temperature and polarization anisotropies
to the large angular scales. We provide different solutions to
deal with the non-Gaussianity of the cross-spectra estimators at large
angular scales.
Working in harmonic space using the cross-spectra allows to get rid of
noise biases and to minimize the residuals systematics effects by
exploiting the cross-correlation between different CMB maps,
e.g. cross-frequency and cross-datasets.
In this sense, the use of cross-spectra allows to access 
the cosmological information encoded in the CMB maps at different
frequencies and to combine different CMB datasets in a more powerful
way with respect to the pixel based or auto-spectra approach.

We present a detailed description of the cross-spectra statistics in \sect{Sec:cross-spectra-stat}. In \sect{Sec:pcl} we describe the $C_\ell$ estimator that we use for the cross-spectra reconstruction and we define the specifications used to generate realistic cross-spectra simulations based on publicly available CMB data. Furthermore, in \sect{subset:parametrization}, we present the formalism to deal with the non-Gaussianity of the cross-spectra $\hat{C}_\ell$ estimators at large angular scales in the case of our realistic simulation settings. 
Based on this formalism, in \sect{Sec:cross-spectra} we construct two types of cross-spectra based likelihoods: in \sect{Sec:single_field} we present an analytical solution based on the parametrization of the $\hat{C}_\ell$ estimator distribution that is useful for the simplest case of a single-field analysis where correlations can be neglected. 
In \sect{Sec:theory_oHL} we then define a more general method that allows to easily deal with a joint temperature and polarization analysis accounting for both correlations between multipoles and modes (TE, TB, EB).
This more general method is based on the extension of the \cite{Hamimeche:2008ai} (H\&L) approach to the large angular scales analysis and it relies on a re-definition of the H\&L variable transformation allowing to approximate the CMB likelihood function by a multivariate Gaussian at low multipoles and for cross-spectra. 
In \sect{Sec:EEonly} we present the likelihood results in the case of a single-field analysis, describing the validation tests and a comparison of the different methods. As the reference single-field we consider the E-modes polarization to constrain the optical depth to reionization parameter $\tau$. 
The results for the general modified H\&L solution (oHL) 
that accounts for the full temperature and polarization analysis including all correlations are described in Sec. \ref{Sec:results_oHLcorr} where we present constraints of the $\tau$, $r$ and $A_s$ parameters. Also, we discuss the optimality of the oHL method with respect to the pixel based likelihood solutions.
Finally in \sect{Sec:conclusions} we present our conclusions.  
We provide in the appendix \sect{app:cross_distrib} the details of the analytic description of the cross-spectra distribution and in \sect{app:crossVSauto} we discuss the comparison of the auto-spectra and cross-spectra statistics.


\section{Cross-spectra statistics}\label{Sec:cross-spectra-stat}

In order to gain some understanding of the underlying statistics, 
we start by presenting the analytical formalism to deal with CMB
cross-spectra on 
the full sky, which will be generalized in \sect{Sec:pcl} in particular to a cut-sky.
We consider the CMB angular power spectrum obtained by combining the harmonic
coefficients $a_{\ell m}$ of two different full-sky maps $(A,B)$, measured with different
noise spectra \Na and \Nb. For a realistic experimental setting the harmonic coefficients are also convolved with the beam functions of the two maps A and B, $b_\ell^A$ and  $b_\ell^B$. The cross-spectra statistics is defined as:
\begin{equation}\label{eq:clcross}
\clAB=\dfrac{1}{2\ell+1} \sum_{\ell=-m}^{m}a_{\ell m}^A  a_{\ell m}^{B\ast} b_\ell^A b_\ell^B.
\end{equation}
In the \eq{eq:clcross} and in the following we make the hypothesis that the noise and the residual systematics are not correlated between the maps so that the cross-spectra are unbiased estimate of the CMB signal.
The cross-spectra distribution is given by (we refer to the Appendix \sect{app:cross_distrib} for the details of this calculation):
\begin{equation}
 \label{eq:cross}
   p^{A \times B}_N (\hat c) = \dfrac{ N^{(N+1)/2} \abs{\hat
       c}^{(N-1)/2}e^{(N \rho \hat c/{z})}
     K_{(N-1)/2}\left( \dfrac{N\abs{\hat c}}{z}\right)} 
   { 2^{(N-1)/2}\sqrt{\pi} \Gamma(N/2) \sqrt{z} (\sa\sb)^{N/2}},  
\end{equation}

where $\hat c=\clAB$, $z=(1-\rho^2)\sa\sb$, $N=2\ell+1$ is the number of modes, $K_\nu$ is the modified
Bessel function of the second kind and order $\nu$ and:   
\begin{equation}
\begin{cases}
  \sa=\sqrt{\clth+N_\ell^A} \\
  \sb=\sqrt{\clth+N_\ell^B} \\
  \rho=\dfrac{\clth}{\sqrt{(\clth+N_\ell^A)(\clth+N_\ell^B)}}.
\end{cases}
\end{equation}

\begin{figure}
  \centering
  \includegraphics[width=\linewidth]{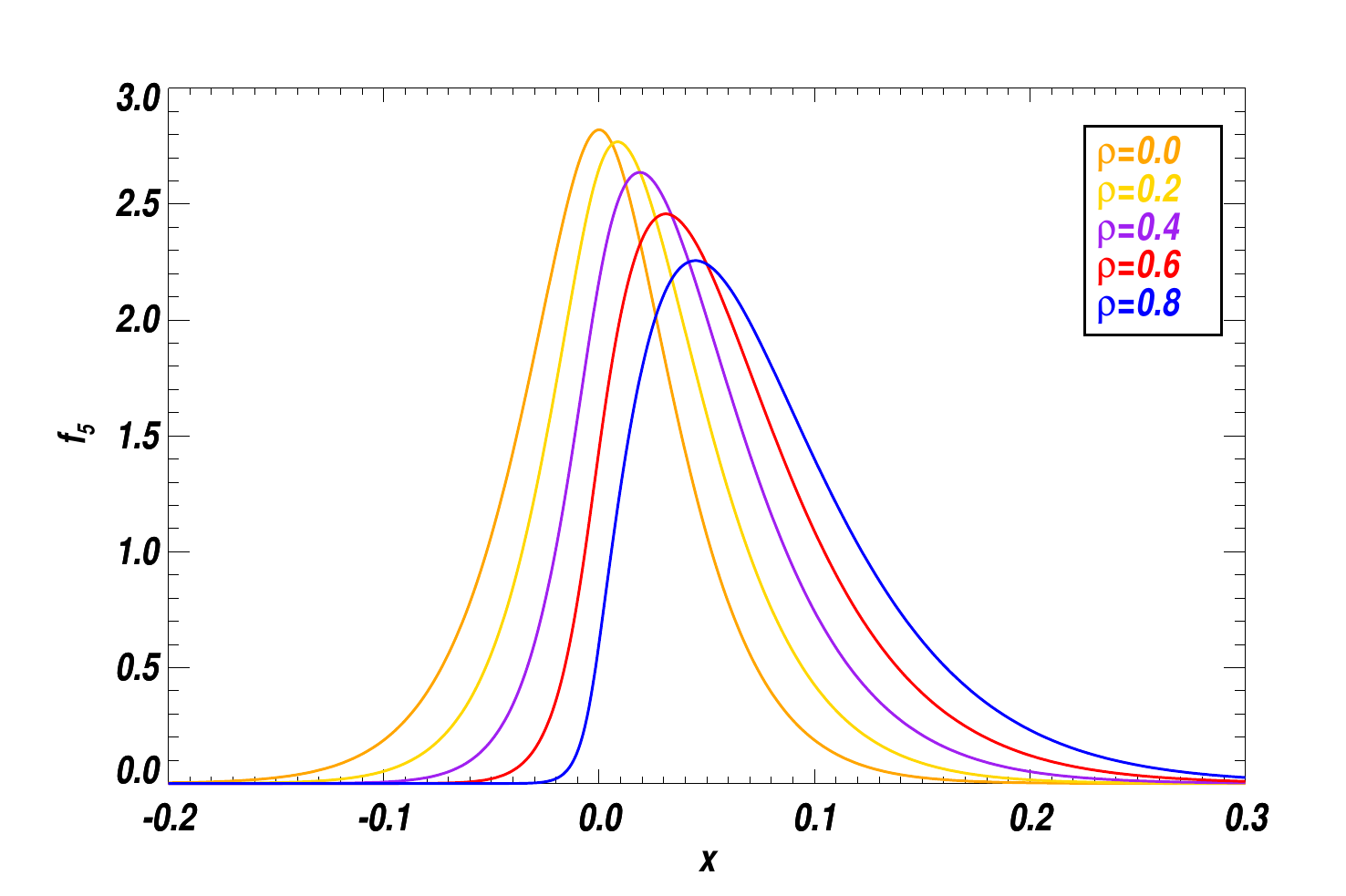}
  \caption{\label{fig:cross_pdf}{Examples of the full-sky cross-spectra
      distributions $p^{A \times B}_N(x\equiv \clAB)$ for $N=5$ modes ($\ell=2$)
      with $\sa\sb=0.1$
      and varying the degree of correlation (\cf \refeq{cross}). }}
\end{figure}

Some examples of the shapes for these distributions are shown on 
Fig. \ref{fig:cross_pdf} where it is interesting to see how the distribution changes when varying the degree of correlation
between the two maps ($\rho$).
Note that, unlike for auto-spectra, the $p^{A \times B}_N (\hat c)$ function can be negative since the
negative exponential decay is compensated by the rise of the
Bessel function, especially when the noise is important (large $\sa,\sb$).

From the characteristic function \refeq{characfunc} we can compute the cumulant generating function
$K(t)=\ln{\phi}(t)$ 
and by Taylor-expanding it in powers of $(it)$ around zero we 
obtain the first cumulants:
\begin{eqnarray}
	\kappa_1(\hat \cl^{A \times B})&=&\clth  \\
	\label{eq:varcross}
	\kappa_2(\hat \cl^{A\times B})&=&\dfrac{2(\clth)^2+\clth(N^A_\ell+N^B_\ell)+ N^A_\ell N^B_\ell}{N} \\
	\label{eq:skewcross}
	\kappa_3(\hat \cl^{A\times B})&=&\clth\dfrac{8(\clth)^2+6\clth(N^A_\ell+N^B_\ell)+6N^A_\ell N^B_\ell}{N^2}.
	\label{eq:meancross}
\end{eqnarray}
This generalize the results from \citet[][Appendix C]{Hamimeche:2008ai}) obtained with identical noise. 
According to the Central Limit Theorem, the cumulants above $\kappa_2$ disappear with the number of modes $N$ and the distribution approaches a Gaussian with a variance given by \refeq{varcross}.
Unlike for the auto-spectrum case (\eq{eq:auto-cumulants}), the estimator $\kappa_1$ does not depend on the noise reconstruction. The clear advantage of using the cross-spectra is that the estimator is unbiased whatever
knowledge we have of the noise spectra. 
Also, the statistical loss for using cross-spectra with respect to auto-spectra is small and minimized if the noise levels of the two maps involved are not too different, as shown in details in  \sect{app:crossVSauto}. 
Note that in general these conclusions hold true also when an incomplete sky coverage is considered. 

\section{Cross-spectra estimator}\label{Sec:pcl}
As in the auto-spectrum case \citeg{Wandelt:2001}, the inclusion of some cut on the sky and of anisotropic noise complicates the
description of a cross-spectrum estimator by correlating modes between them (both in $\ell$ and $m$ for a non azimuthal mask) and eventually distorting the marginal distributions.
We then need to rely on realistic simulations to take into account the full complexity of the problem.

\subsection{Angular power spectrum estimator}

\begin{figure}
	\centering
	\includegraphics[width=\columnwidth,height=200px]{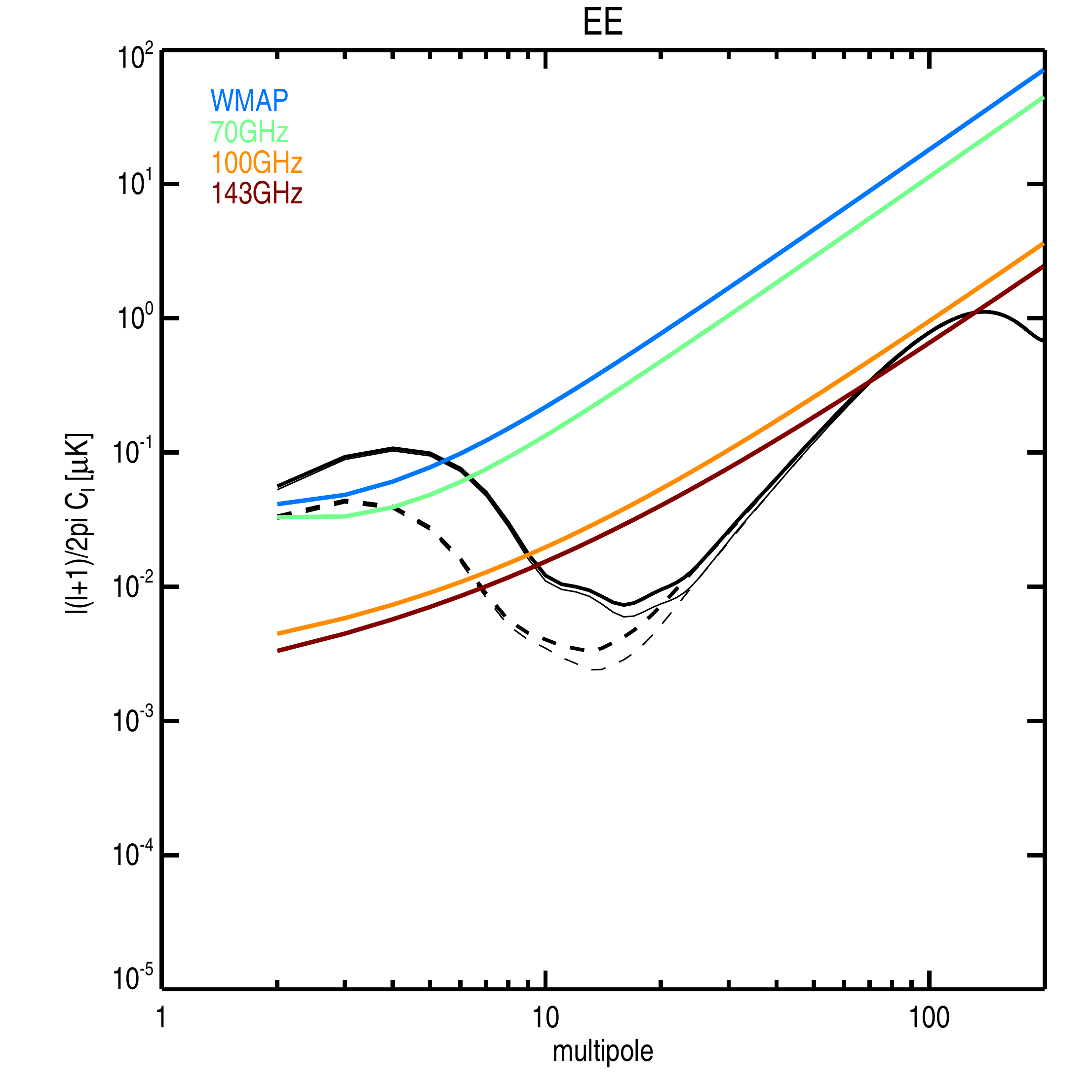}
	\includegraphics[width=\columnwidth,height=200px]{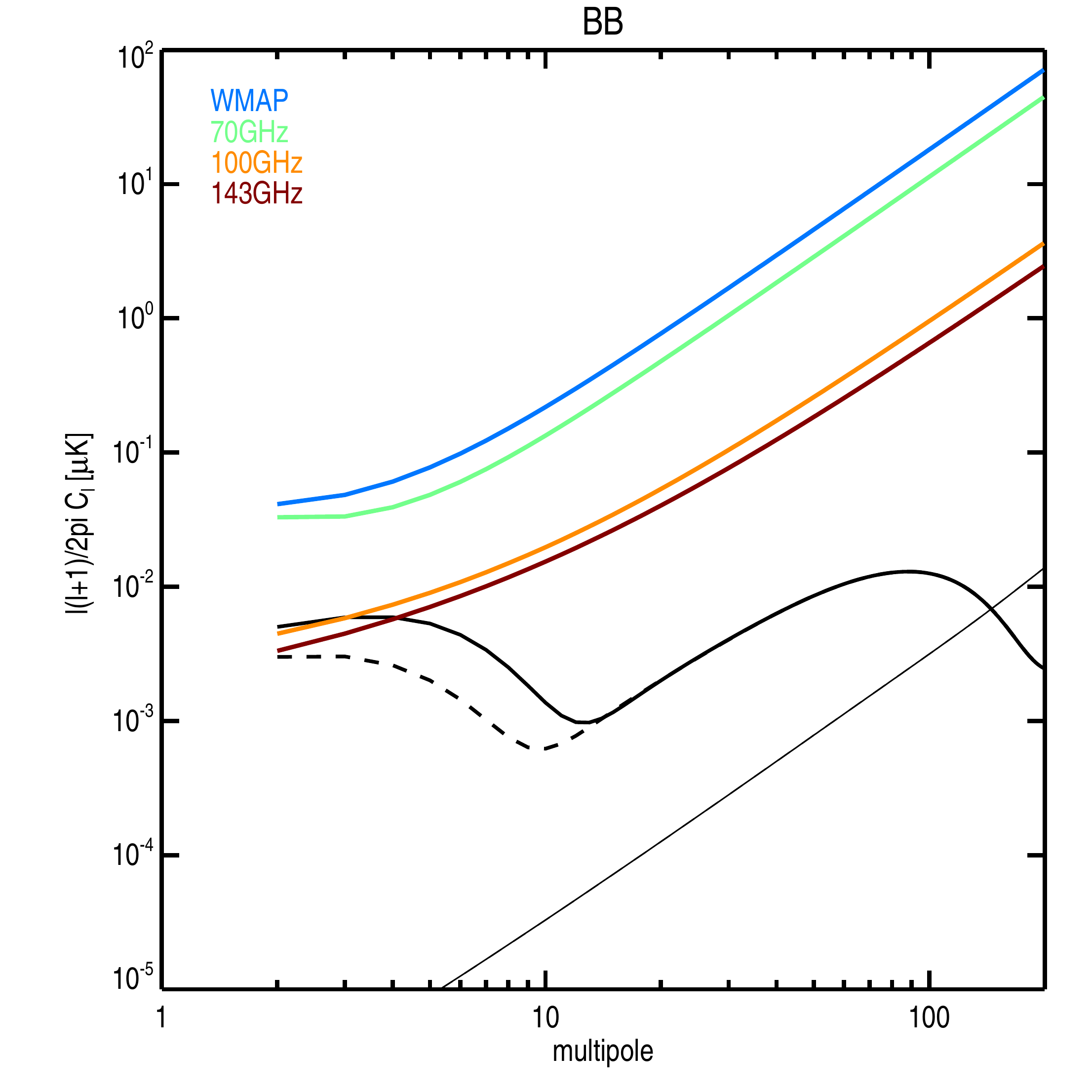}
	\caption{Polarized power spectra for E modes ({\it top}) and B modes ({\it bottom} for different reionization histories: late ($\tau=0.056$, dashed line) or early ($\tau=0.09$, solid line). 
	The primordial B-modes spectra are shown for r = 0.2 (solid and dashed thick lines) and the lensing contribution to the B-modes signal is shown as the thin line.  
	The noise levels for the four considered data cases are over plotted (\wmap, \lfi, \planck-100 and \planck-143).}
	\label{fig:models}
\end{figure}

\begin{figure*}%
	\centering
	\includegraphics[width=\columnwidth]{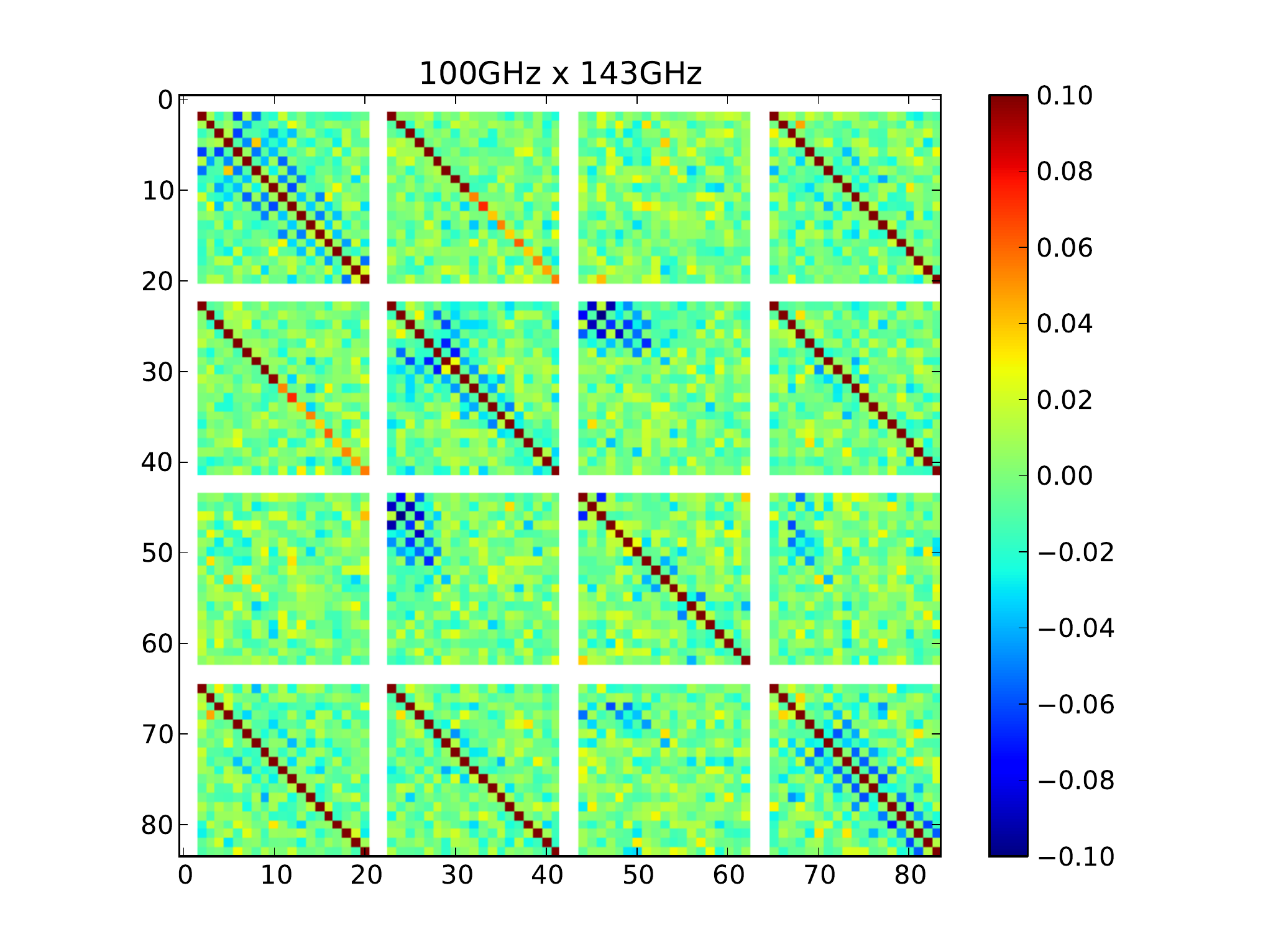}
	\includegraphics[width=\columnwidth]{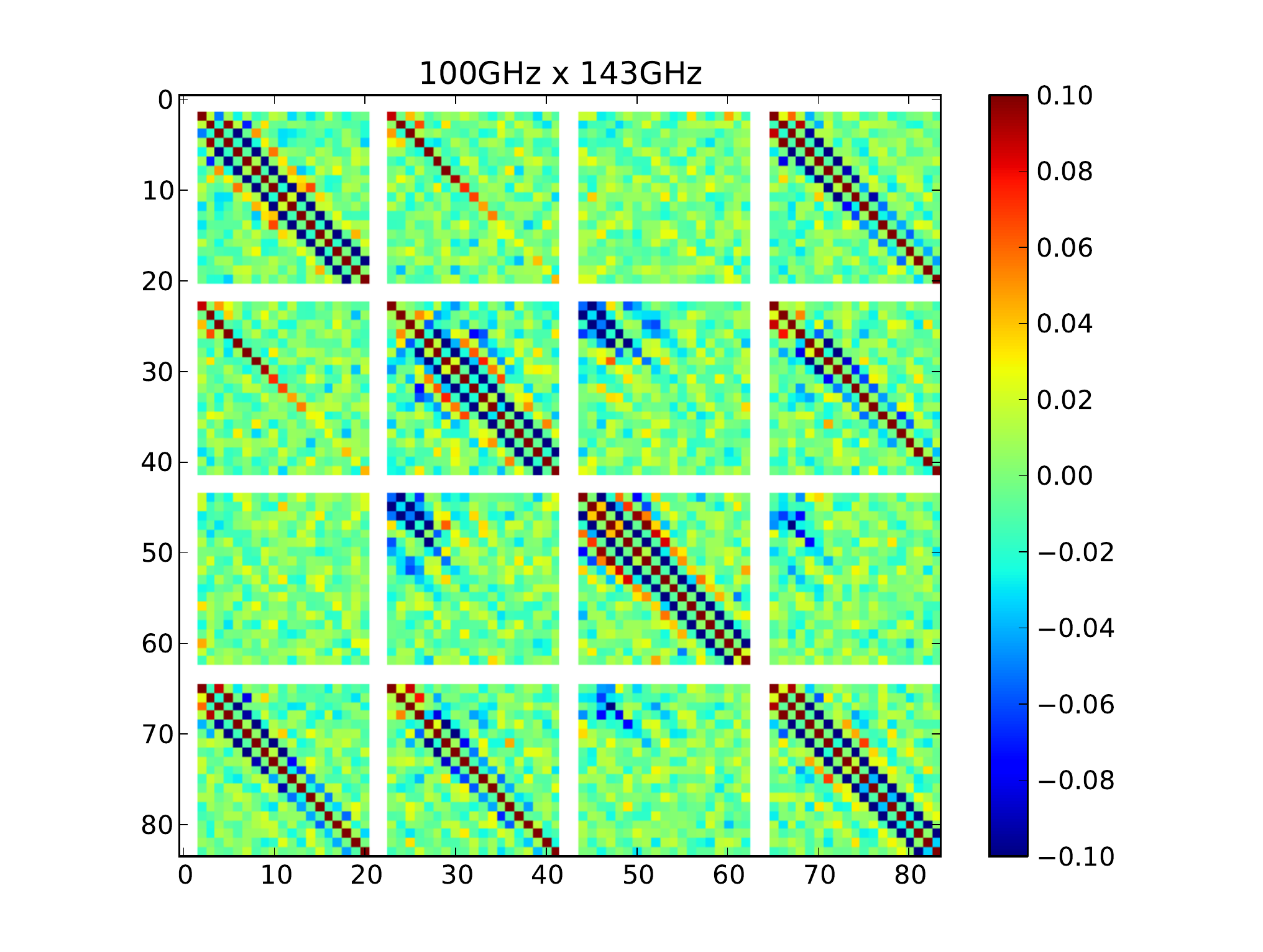}
	\caption{Correlation matrix for the cross-power spectra 100x143 with two different sky coverage: 80\% ({\it left}) and 50\% ({\it right}). Each block corresponds to $TT$, $EE$, $BB$, $TE$ spectra respectively. $\ell=0$ and $\ell=1$ are not defined and set to zero.}
	\label{fig:corrmat}
\end{figure*}

To derive different sets of cross-spectra simulations we use \texttt{Xpol}, a pseudo-$C_\ell$ estimator (PCL) based on the extension to polarization of the \texttt{Xspect} algorithm \citep{Tristram:2004if} .
At very low multipoles, the PCL estimator is known to be sub-optimal with respect to e.g. a Quadratic Maximum Likelihood estimator (QML) \citep{Tegmark:2001,Efstathiou:2006}. This means that the variance and correlation of the PCL is expected to be slightly higher than for QML resulting in slightly larger distributions for the estimated $\hat{C}_\ell$. 
However, the implementation of cross-spectra for PCL estimators is straightforward and we can easily take into account the level of $\ell$-by-$\ell$ correlations using Monte Carlo simulations. In any case, the definition and validation of the cross-spectra likelihood are independent on the choice of the cross-spectra estimator used.

We produced different sets of Monte-Carlo cross-spectra simulations. We generated simulated CMB maps on which we add anisotropic and correlated noise corresponding to four public datasets: WMAP~(V band), \planck-LFI~(70\,GHz) and two \hfi\ channels~(100\,GHz and 143\,GHz).
The CMB signal is constructed from the adiabatic $\Lambda$CDM model with cosmological parameters: $\Omega_b h^2$ (the baryon density), $\Omega_c h^2$ (the dark matter density), the amplitude and the spectral index of the primordial power spectrum $A_s$ and $n_s$, $\theta$ (a parameter proportional to the ratio of the sound horizon and the angular diameter distance at recombination), the optical depth to reionization parameter $\tau$ or, equivalently, the redshift of reionization $z_{re}$. %
We also consider primordial tensor modes, parametrized by the tensor-to-scalar ratio of the amplitude of the primordial spectra $r$. 

Our reference simulations are generated with a fiducial $\Lambda$CDM model based on the \planck\ 2015 best fit \citep{planck2015-XIII} with $\tau=\tauvp$. For the tensor-to-scalar ratio we choose $r=0.1$ as the fiducial input value.
Since it is relevant for some validation tests, in particular to check the dependence on the fiducial model, we also generated two sets of simulations with different input cosmologies: 
\begin{enumerate}[model 1:]
\item early reionization without tensor modes (\planck\ 2015 best-fit with $\tau=0.09$, $z_{re}=11.2$, $r=0$)
\item late reionization with high level of tensor (\planck\ 2015 best-fit with $\tau=0.0566$, $z_{re}=8$, $r=0.2$)
\end{enumerate}

We estimate the noise angular power spectrum in temperature and polarization using the spectra of ($I$,$Q$,$U$) year map differences for \wmap\ and \lfi. For \hfi, we compute the temperature spectrum from available HFI intensity year map differences which we rescale according to the number of polarized detectors at each frequency to mimic the polarized noise power spectra. The latter ends up very close to what is published in \cite{planck2015-VIII}. 
With this procedure, the noise power spectra used for the simulations include realistic white noise level and low-frequency noise due to systematic and foreground residuals.
From those power spectra, we derive constrained map realization of noise for each simulation. We then scale the noise map by the appropriate relative hit counts in each pixel to simulate the inhomogeneous scanning of each dataset.

We use our pseudo-$C_\ell$ estimator to produce the six cross-spectra corresponding to the four datasets: \wmap$\times$\lfi, \wmap$\times$\hfic, \wmap$\times$\hficq, \lfi$\times$\hfic, \lfi$\times$\hficq\ and \hfic$\times$\hficq. For each simulation, we construct the $TT$, $EE$, $BB$, $TE$, $TB$ and $EB$ cross-power spectra.
The upper and lower panels of \fig{fig:models} show the different noise levels corresponding to the four datasets for the E-modes and B-modes spectra, respectively and how the CMB polarized power at very low multipoles is directly scaled by the optical depth of reionization. 

The plots illustrate the effect of the change form early ($z_{re} = 11.2$) to very late ($z_{re}=8$) reionization -- which correspond to an optical depth of $\tau=0.09$ and 0.566 respectively -- for both E and B modes below $\ell=10$. In addition, the tensor-to-scalar ratio rescales the overall amplitude of the primordial signal in BB. We always include the lensing contribution to the B-modes shown as the thin line in the lower panel of \fig{fig:models}.

We do not simulate the impact of foreground contaminations in map domain. However, residuals from foreground contaminations are statistically included in our estimate of the noise spectra. Moreover, we remove the Galactic plane for the power spectrum estimation. We use two sets of Galactic mask based on a threshold on the polarized power amplitude of the dust emission and allowing for a sky coverage of 80\% and 50\% respectively.

The correlation matrices (Fig.~\ref{fig:corrmat}) are directly derived from the Monte Carlo (MC). The level of correlation between multipoles depend on the sky cut and the dataset considered. For the \planck-100x\planck-143, using 80\% sky coverage the correlations are weak, lower than 5\%, as shown in the left panel of Fig.~\ref{fig:corrmat}. As we will see in the next section (Sect.~\ref{subset:parametrization}), for such a large sky coverage we can safely neglect the correlations and adapt the full-sky cross-spectra statistic. For the 50\% sky, the correlations are significantly higher and can reach the level of 25\% (right panel of Fig.~\ref{fig:corrmat}).

\subsection{Parametrization of the PCL marginals}\label{subset:parametrization}

The distribution of the PCL estimator is largely non-gaussian in the low-$\ell$ regime we are interested in (see examples in \fig{fig:distrib5_anal} and \fig{fig:distrib5_edge} in the Appendix \ref{app:fit_distribs}) and all order moments actually depend on the noise powers and on the fiducial model. Leaving aside the complicated (and unnecessary) task of defining the full joint \pdf $p(\vec \cl)$, we focus on how to parametrize analytically the individual (\ie marginal) distributions by tweaking the results obtained on the full-sky in \sect{Sec:cross-spectra-stat} in the case that a sky cut is applied. We propose two different approaches to achieve a satisfactory description.

\subsubsection{Full-sky based approach}\label{subsec:result_analytic}
A somewhat heuristic argument used when masking some fraction \fsky of the sky, is to consider that the \q{number of degrees of freedom} of the associated $\chi^2$ distribution $N=(2\ell+1)$ is reduced asymptotically by the \fsky factor \citep{Hivon2002}. When the mask is apodized by some window, we include the weighting factor ${w_2^2}/{w_4}$, where $w_i$ is the i-th moment of the weighting scheme, in our definition of \fsky.
  
Keeping in mind that cross-spectra do not follow any $\chi^2$ distribution
and that we are not in the asymptotic regime, 
we may still try to adapt this methodology based on our simulations.
We then modify our number of modes by $N=(2\ell+1)\fskyc$, keep
the general full-sky shape of \refeq{cross} and fit for the \fskyc
factor for different masks, noise combinations and models.

  \begin{figure}
  \centering
  \begin{tabular}{c}
 \includegraphics[width=\linewidth]{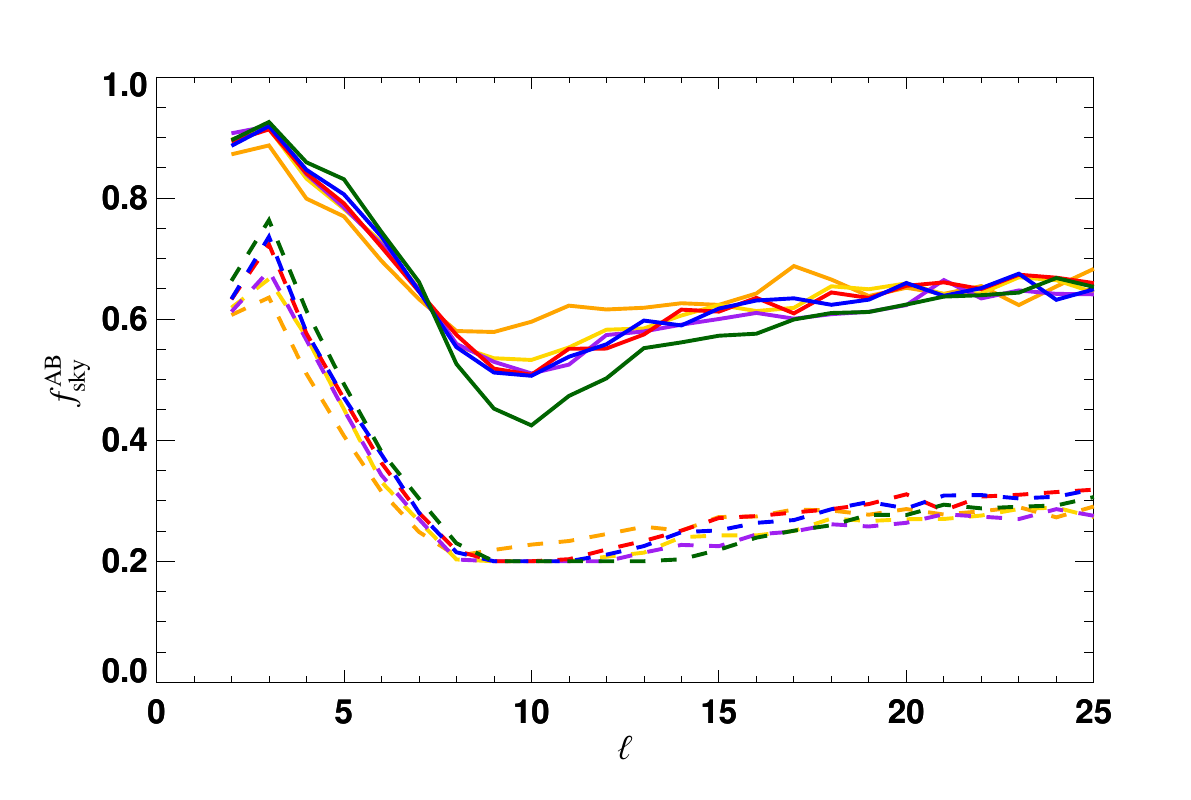} \\
 \includegraphics[width=\linewidth]{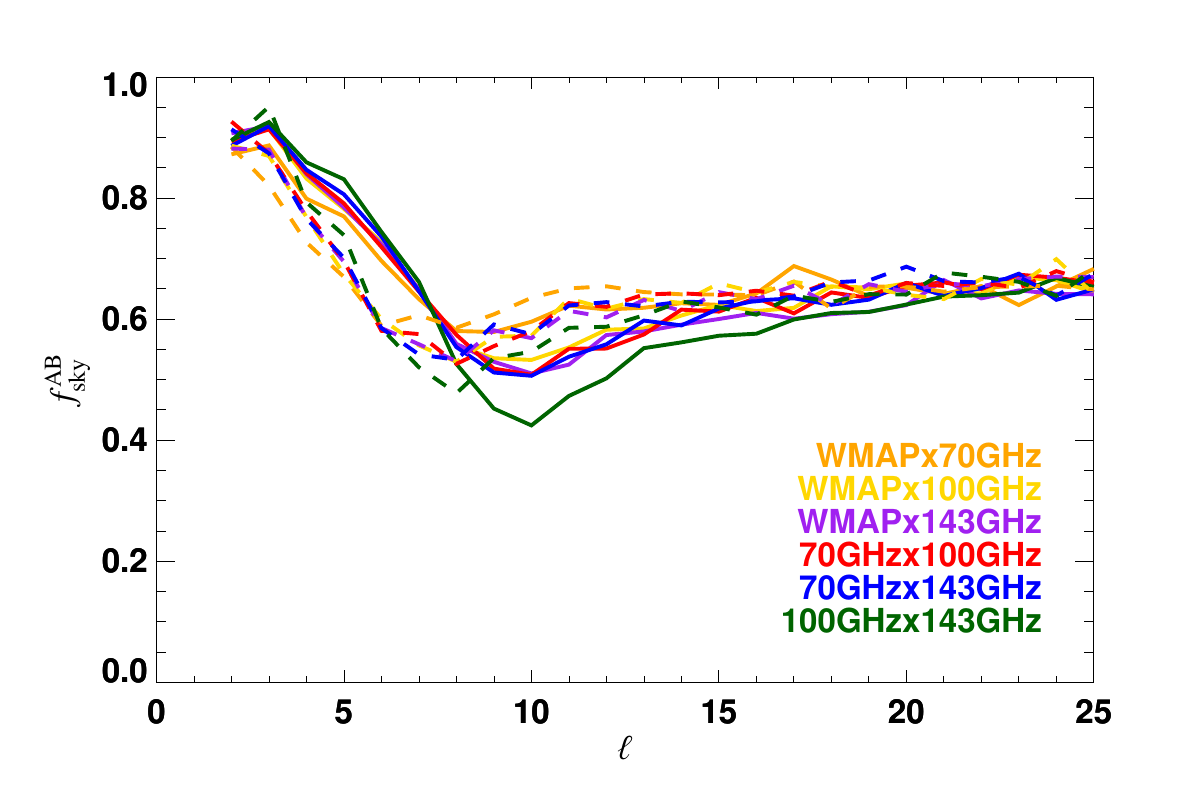}
 \end{tabular}
  \caption{
  \label{fig:fsky_pc} 
  Upper panel: {\it Dependence of the \fskyc factor on the mask.}
  The two distinct groups (full and dashed lines) represents the
  \fskyc values fitted to the distributions  obtained
  respectively on the small mask (\fsky=0.77)
  and large (\fsky=0.49) mask. 
 Lower panel: {\it Dependence of the \fskyc factor on the fiducial model.}
The plot shows the \fskyc factor as a function of the multipole when the early-reionization model (model 1, $\tau=0.09$, solid lines) and the late-reionization model (model 2, $\tau=0.056$, dashed lines) are used as input cosmology in the simulations.} 
\end{figure}

The upper panel of Fig. \ref{fig:fsky_pc} shows the $\fskyc$ factor as a function of the
multipole $\ell$ in the case of cross-spectra with different noise
levels for the small (\fsky=0.77) and larger (\fsky=0.49) mask.
As expected, there is a strong dependency of $\fskyc$ on the mask
size.
The $\fskyc$ is not a constant in the low-$\ell$ regime
($\lesssim 15$) and it is asymptotically slightly different from the
standard \fsky factor. This can be traced to the fact that we are
dealing with polarization that involves different Wigner 3j functions
than the ones derived from temperature. 
Despite this strong dependence on the mask, the $\fskyc$
functions derived for the six cross-spectra show a very good
consistency for all those different noise levels.

In the lower panel of \fig{fig:fsky_pc} we check the model dependency by
comparing the $\fskyc$ values reconstructed from two simulation sets generated with two different input cosmologies: the early reionization scenario (model 1, $\tau=0.09$) and the late reionization scenario (model 2, $\tau=0.056$).
The reconstruction of the \fskyc factor is reasonably stable with
respect to the change in the fiducial model used and will be
considered in the following as independent.

The stability with respect to the choice of the fiducial model is further demonstrated in the Appendix \ref{app:fit_distribs}
(Fig.~\ref{fig:distrib5_anal}) which shows the excellent agreement of
the \pdf's for the \planck-100$\times$\planck-143 cross-spectrum estimator 
\textit{for both models} while $\fskyc(\ell)$ is derived from \textit{a single one}.
Note that we choose to display our \q{worse} case: all other
cross-spectra parametrisations are even better.

\subsubsection{Edgeworth expansion}\label{subsec:results_edge}

As an alternative to the reconstruction of the $\fskyc(\ell)$ function,
we also propose another approach, noticing that only the first
three central moments contribute essentially to the estimator distribution
above $\ell \gtrsim 4$. In this case we use the standard (constant) \fsky factor in $N=(2\ell+1)\fsky$ and, to account for the fact that a fraction of the sky is masked, we modify the coefficients of the \clth-polynomial
of the full-sky cumulants
(Eqs.~\ref{eq:varcross} and \ref{eq:skewcross}) to match the ones
reconstructed from the simulations. The new cumulants in this more general case take the
form:

\begin{align}
\label{eq:cucu}
  \kappa_1(\hat \cl^{A\times B})&=\clth \nn \\
  \kappa_2(\hat \cl^{A\times B})&=\dfrac{1.5(\clth)^2+2\clth(N^A_\ell+N^B_\ell)+ N^A_\ell
    N^B_\ell}{N} \\
  \kappa_3(\hat \cl^{A\times
    B})&=\clth\dfrac{6(\clth)^2+12\clth(N^A_\ell+N^B_\ell)+ 10N^A_\ell
    N^B_\ell}{N^2}. \nn
\end{align}

\begin{figure}
	\centering
	\includegraphics[width=\linewidth]{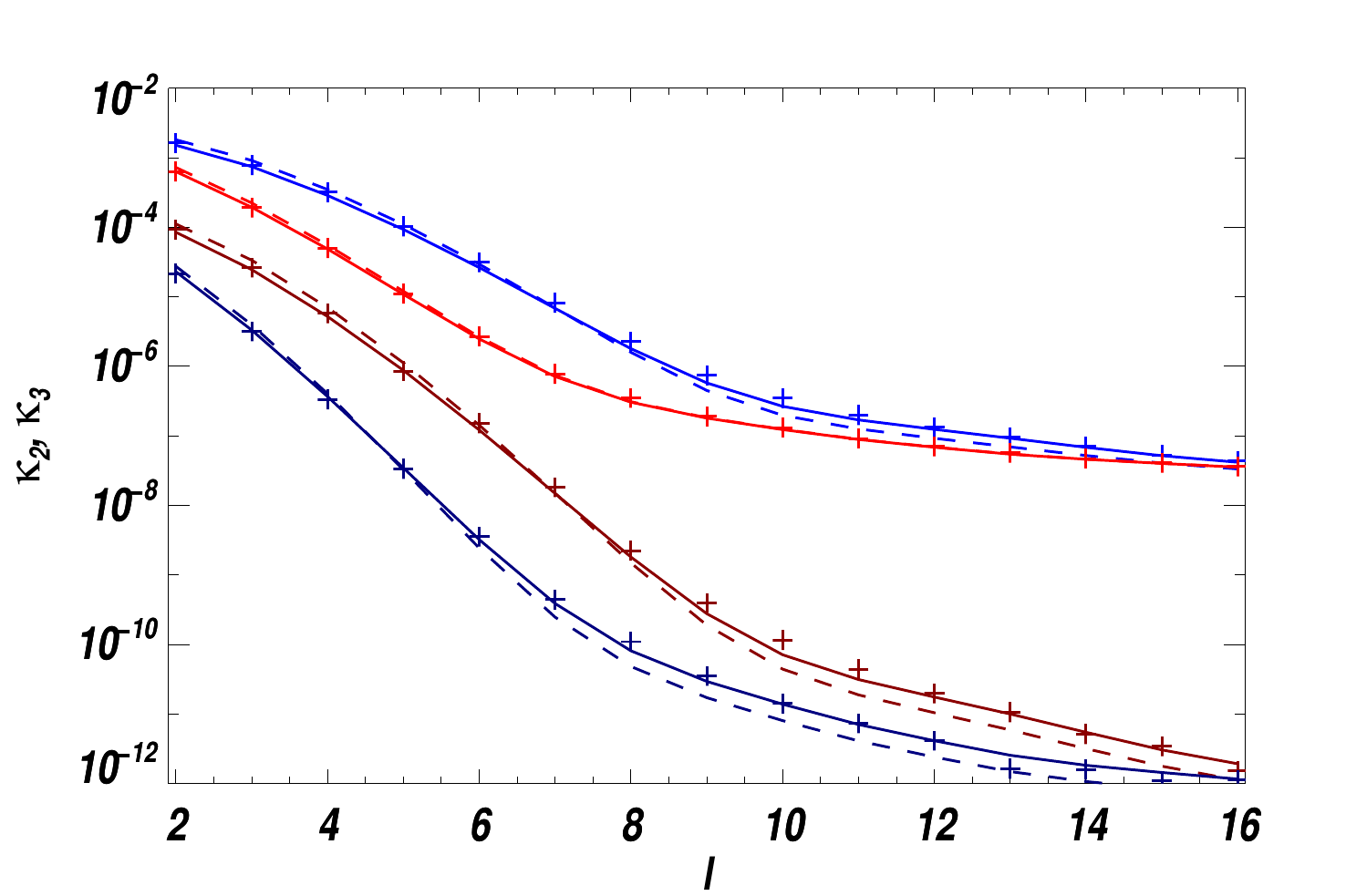}
	\caption{\label{fig:kumul23}
{\it Variance and skewness of the cross-spectra estimators.} The plot shows the second order cumulants, $\kappa_2$ (upper red and blue curves) and third order $\kappa_3$ (lower red and blue curves) of the PCL
estimator. The blue lines correspond to results with the early reionization scenario (model 1) as fiducial model, the red when simulations has the late reionization scenario as input cosmology (model 2). The points refer to the cumulants reconstructed from the \hfic$\times$\hficq\ simulations.
The dashed lines correspond to the analytic full-sky derived
expressions Eqs.~\ref{eq:varcross}--\ref{eq:skewcross} with a rescaled $N=(2\ell+1)\fsky$ factor. The solid lines refer to
the parametrization of the cumulants based on the Edgeworth expansion defined in \refeq{cucu}.}
\end{figure}

Note that this parametrization just depends on constant values of the polynomial
coefficients. Fig.~\ref{fig:kumul23} shows the $\kappa_2$ (variance) and $\kappa_3$ (skewness) reconstructed from the simulations for the two input fiducial models (early reionization model in blue and late reionization model in red). The agreement between the full-sky based approximation (dashed lines) and the parametrization of \refeq{cucu} derived from simulation (solid lines) is excellent. We emphasize that the $\kappa_3$ tuning is not mandatory (one may use the one from \refeq{skewcross}) since it drops rapidly.

The optimization of the cumulants was performed on a single
cross-spectrum (\planck-100$\times$\planck-143, model 2).
It is however robust enough to be used in all other cases as will be
demonstrated later in the likelihood tests (\sect{Sec:EEonly}).

Now that we have a model-independent parametrization of the first cumulants, we
proceed in writing an analytical description of the estimator \pdf
using an Edgeworth Series expansion \citeg{Kendall}. Using the
normalized variable $y=\dfrac{\hat \cl-\mu}{\sigma}$ where
$\mu=\kappa_1$ and $\sigma=\sqrt{\kappa_2}$, the truncated expansion reads:
\begin{equation}
  \label{eq:edge}
  f(y|\clth,\Na,\Nb)=\mathcal{N}(y)\left(1+\dfrac{\kappa_3}{6\sigma^3}
    H_3(y)\right),
\end{equation}
where $\mathcal{N}$ denotes the normal distribution and $H_3$ is the
3rd order \q{probabilistic} Hermite polynomial \citep{Kendall}. Each
$\mu,\sigma,\kappa_3$ is computed from \refeq{cucu} and depends only
on $\clth,\Na,\Nb$.

A classical issue when truncating an Edgeworth expansion is
that, despite being properly normalized to one, it may lead to negative
values so that \refeq{edge} is not really a \pdf\ and may lead to
potential problems when constructing with it a log-likelihood  function. We adopt the method proposed by \citet{Rocha2001} to alleviate this
problem. Their idea is based on 
the solutions of the quantum harmonic oscillator, that exhibits
non-Gaussianity above the ground level. 
For one extra-level the wave-function (\ie a \pdf) is of the form:
\begin{equation}
\label{eq:edgepp}
  P(x)=\mathcal{N}(x)\left( \alpha_0+\dfrac{\alpha_3}{\sqrt{6}} H_3(x) \right)^2,
\end{equation}
with $\alpha_0=\sqrt{1-\alpha_3^3}$.
For a mild non-Gaussianity (small $\alpha_3$), which is the case in our regime,
we expand this equation:
\begin{equation}
  P(x)=\mathcal{N}(x)\left(1
    +\dfrac{2\alpha_3}{\sqrt{6}}H_3(x)+O(\alpha_3^2) \right),
\end{equation}
and equating terms to \refeq{edge}, we identify:
\begin{equation}
  \label{eq:alpha3}
  \alpha_3=\dfrac{\kappa_3}{2\sqrt{6}\sigma^3}.
\end{equation}

In the following we will refer to \q{Edgeworth expansion} as this regularized form, namely \refeq{edgepp} using \refeq{alpha3}.

Fig.~\ref{fig:distrib5_edge} shows the agreement between the empirical estimator distributions obtained on simulations and our Edgeworth-based parametrization.
The agreement is very satisfactory but for the $\ell=2,3$ case which
would require the use of higher order cumulants. On the other side,
introducing some $\kappa_4$ (kurtosis) term brings some oscillations
upon all the multipoles. This would not be desirable since the first
two accessible multipoles have generally a very low SNR due to $1/f$
noise and large cosmic-variance and can be disregarded without a sizable loss of information.

\section{Cross spectra-based likelihoods}\label{Sec:cross-spectra}
With these tools in hand we now proceed in constructing the likelihood of a given model, which means inverting the (unknown) joint and possibly multi-field PCL estimator distribution given the true value \clth. 
We first discuss the simple but frequent single-field case with a small mask for which we give analytical formulas.
We then define a more general solution based on the modification of the H\&L approximation to construct a general likelihood solution for the combination of the temperature and polarization fields accounting for correlations.

\subsection{Single field approximations neglecting correlations}\label{Sec:single_field} 

As a first solution, we can build our real-case likelihood from the parametrization of the marginalized estimator distribution $p(\cl)$ described in \sect{subset:parametrization}.
This approximation is accurate when the masked sky-fraction is low
(typically below $20\%$) so that the $\ell$-by-$\ell$ correlations can be safely neglected.
The likelihood function is defined as the product of the probability density functions $p_N^{A\times B}$  (cfr. \eq{eq:cross}):

\begin{equation}
   \label{eq:lik_anal}
\lik^{A\times B}(\clth|\hat \cl,\Na,\Nb)=\prod_{\ell=\lmin}^{\lmax} p_N^{A\times B}(\hat \cl),
\end{equation}
where the $\hat \cl$ represent the values measured on data. The $p_N^{A\times B}$ functions are implicitly dependent on $\Na$, $\Nb$, $\clth$ and they can be defined according to the chosen analytical parametrization as described in \sect{subsec:result_analytic} and \sect{subsec:results_edge}.

Note that this approximation derived from the full sky formalism is easily defined for a single field, that is to say when the cross-spectra are derived from the combination of the same temperature or polarization field, e.g. the E-modes cross-spectra.
A combined analysis that accounts for all the temperature and polarization fields is difficult to define analytically since correlations between different fields (TE, TB, EB) cannot be neglected and higher order moments of the $\hat \cl$ distribution must be accounted, making the analytical solution difficult to handle in this more general case.

\subsection{General multi-field approximation}\label{Sec:theory_oHL}

Here we present a more general formalism to define a cross-spectra likelihood for the analysis of the CMB data at large angular scales that allows to deal with realistic cases of incomplete sky coverage taking into account the $\ell-\ell$ correlations. This likelihood can also be easily generalized to a multi-fields likelihood that combines the temperature and polarization fields T, E and B, accounting for the field-field correlations TE, TB and EB.

In order to model the non-Gaussianity of the $\hat \cl$ estimators, the approximation that we propose is based on the modification of the Hamimeche\&Lewis likelihood (H\&L) \citep{Hamimeche:2008ai}, adapted to work for the cross-spectra $C^{A\times B}_{\ell}$ and at low-$\ell$.

The general form of the H\&L likelihood is defined for auto-spectra at intermediate and small scales ($\ell>30$) \citep{Hamimeche:2008ai}:
\begin{equation}
-2ln\lik(C^{th}_\ell|\hat{C}_\ell)=\sum_{\ell \ell'} [X_g]^T_\ell [M_f^{-1}]_{\ell \ell'} [X_g]_{\ell'}.
\end{equation}
The $[M_f^{-1}]_{\ell \ell'}$ is the inverse of the $C_\ell$-covariance matrix that allows to quantify the $\ell-\ell$ and the correlations of the T, E, B fields. 
The vector $[X_g]_\ell$ is the H\&L transformed $C_\ell$ vector defined as:
\begin{equation}
[X_g]_\ell=\mbox{vecp}\Big( {\bf C}^{1/2}_{fid} {\bf U} ({\bf g} [ {\bf D(P)}] ){\bf U^T } {\bf C}^{1/2}_{fid}  \Big).
\label{eq:Xg}
\end{equation}

In eq.\ref{eq:Xg}, ${\bf C}^{1/2}_{fid}$ is the square root of the $C_\ell$ matrix:
\begin{center}
\begin{equation}
{\bf C}_{\ell} = 
\begin{pmatrix}
 C_\ell^{TT} &C_\ell^{TE}  & C_\ell^{TB} \\
 C_\ell^{TE} &C_\ell^{EE} & C_\ell^{EB} \\
 C_\ell^{TB} &C_\ell^{EB}  & C_\ell^{BB}
\end{pmatrix}
\label{eq:clmatrix}
\end{equation}
\end{center}
for a given fiducial model and the function $({\bf g} [ {\bf D(P)}] )$ refers to the transformation: 
\begin{equation}
g(x)=sign(x-1) \sqrt{(2(x-ln(x)-1))},
\label{eq:ghl}
\end{equation}
applied to the eigenvalues of the matrix ${\bf P}={\bf C}^{-1/2}_{mod} {\bf \hat{C}}_{data}  {\bf C}^{-1/2}_{mod} $, where ${\bf C}_{mod}$ and  ${\bf \hat{C}}_{data}$ are, respectively, the matrices of the sampled $C_\ell$ and the data. 
This approximation has been shown to be robust with respect to the choice of the fiducial model \citep{Hamimeche:2008ai}.
The problem is that for cross-spectra and at large angular scales 
the ${\bf P}$ matrix is no longer guaranteed to be positive definite. In fact, as shown in \eq{eq:cross} that describes the distribution of the cross-spectra estimators $\hat \cl$, the ${\bf \hat{C}}_{data}$ can be negative. 
In order to solve for this issue, we propose a modification of the H\&L likelihood that consists in adding an effective offset $o_\ell$ to the cross-spectra. 
This mimics the noise bias of the auto-spectra and makes the offset-cross-spectra distribution very similar to the auto-spectra distribution used in the H\&L approximation.
We re-define each ${\bf C}_{\ell} $ matrix (eq. \ref{eq:clmatrix}) as:
\bea
{\bf C}^{A\times B}_{\ell} \rightarrow O({\bf C}^{A\times B}_{\ell} )= 
\begin{pmatrix}
 C_\ell^{TT} + o^{TT}_\ell&C_\ell^{TE}  & C_\ell^{TB} \\
 C_\ell^{TE} &C_\ell^{EE}+o^{EE}_\ell & C_\ell^{EB} \\
 C_\ell^{TB} &C_\ell^{EB}  & C_\ell^{BB}+o_\ell^{BB}	
\end{pmatrix}
\label{eq:oHLgeneral}
\eea
so that:
\begin{equation}
[X_g({\bf C}^{A\times B}_{\ell} )]_\ell \rightarrow [OX_{g}]_\ell= [X_{g}(O({\bf C}^{A\times B}_{\ell} ))]_\ell.
\label{eq:oHLtrasf}
\end{equation}

The new offset H\&L likelihood (oHL hereafter) reads:
\begin{equation}
-2ln\lik(C_\ell|\hat{C}^{A\times B}_\ell)=\sum_{\ell \ell'} [OX_g]^T_\ell [M_f^{-1}]_{\ell \ell'} [OX_g]_{\ell'}.
\label{eq:oHL}
\end{equation}
The variable transformation $g(x)$ is now modified for the cross-spectra to regularize the likelihood around zero so that \eq{eq:ghl} now reads:
\begin{equation}
g(x) \rightarrow sign(x) g(\vert x \rvert).
\label{eq:gcross}
\end{equation}

The offset function $o_\ell^{XY}$ can be derived from simulations.  
We estimate the offsets from the MC distributions ensuring that the {\bf P} matrix reconstructed is positive definite for more than 99\% of our simulations. 
\begin{figure}
\centering
\includegraphics[width=\linewidth]{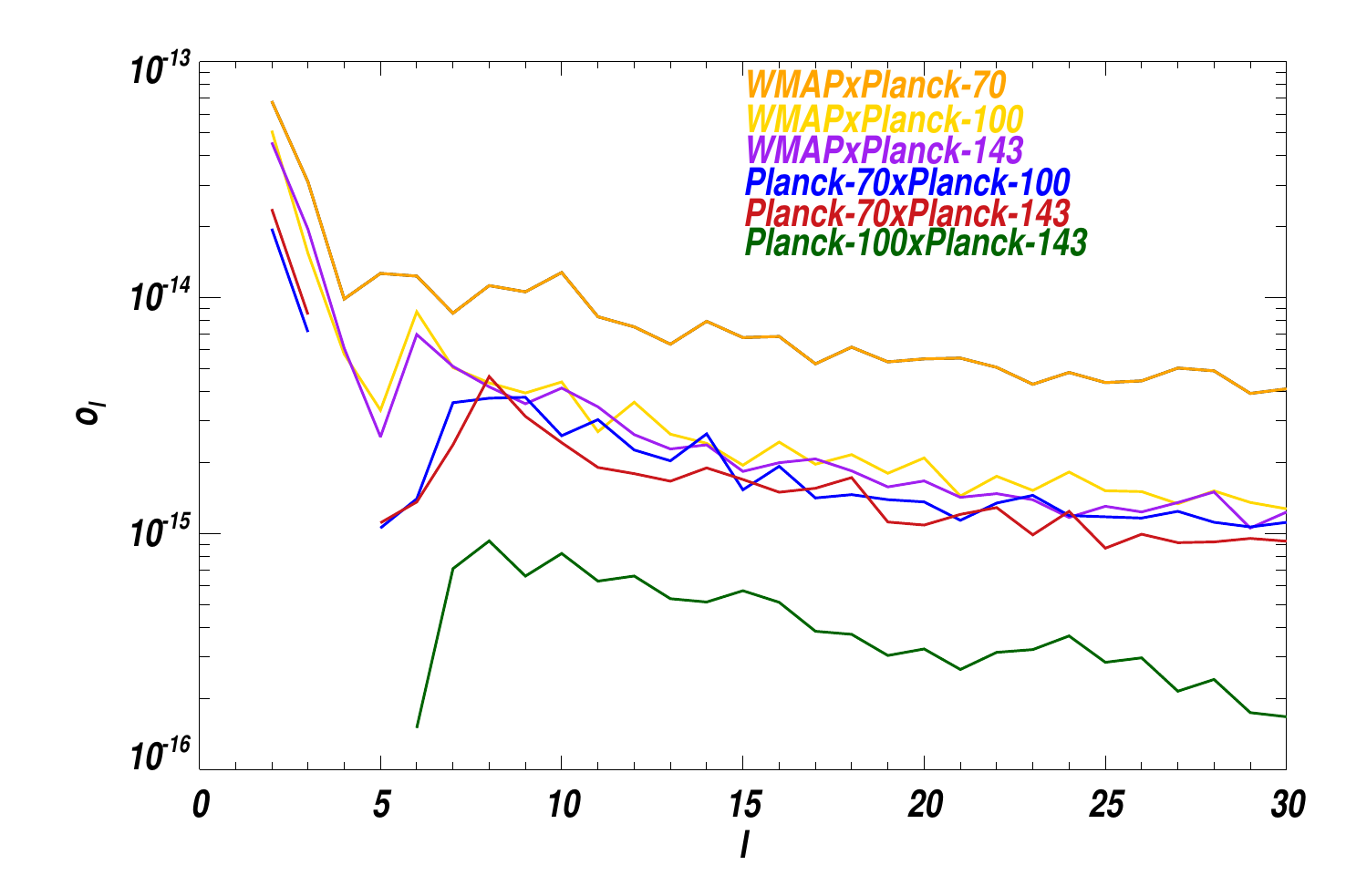}
\caption{ The offset functions $o^{A\times B}_\ell$ of the oHL likelihood for the E-modes cross-spectra and for the six different combinations of noise levels.}
\label{Fig:offsetEE}
\end{figure}
\begin{figure}
\centering
\includegraphics[width=\linewidth]{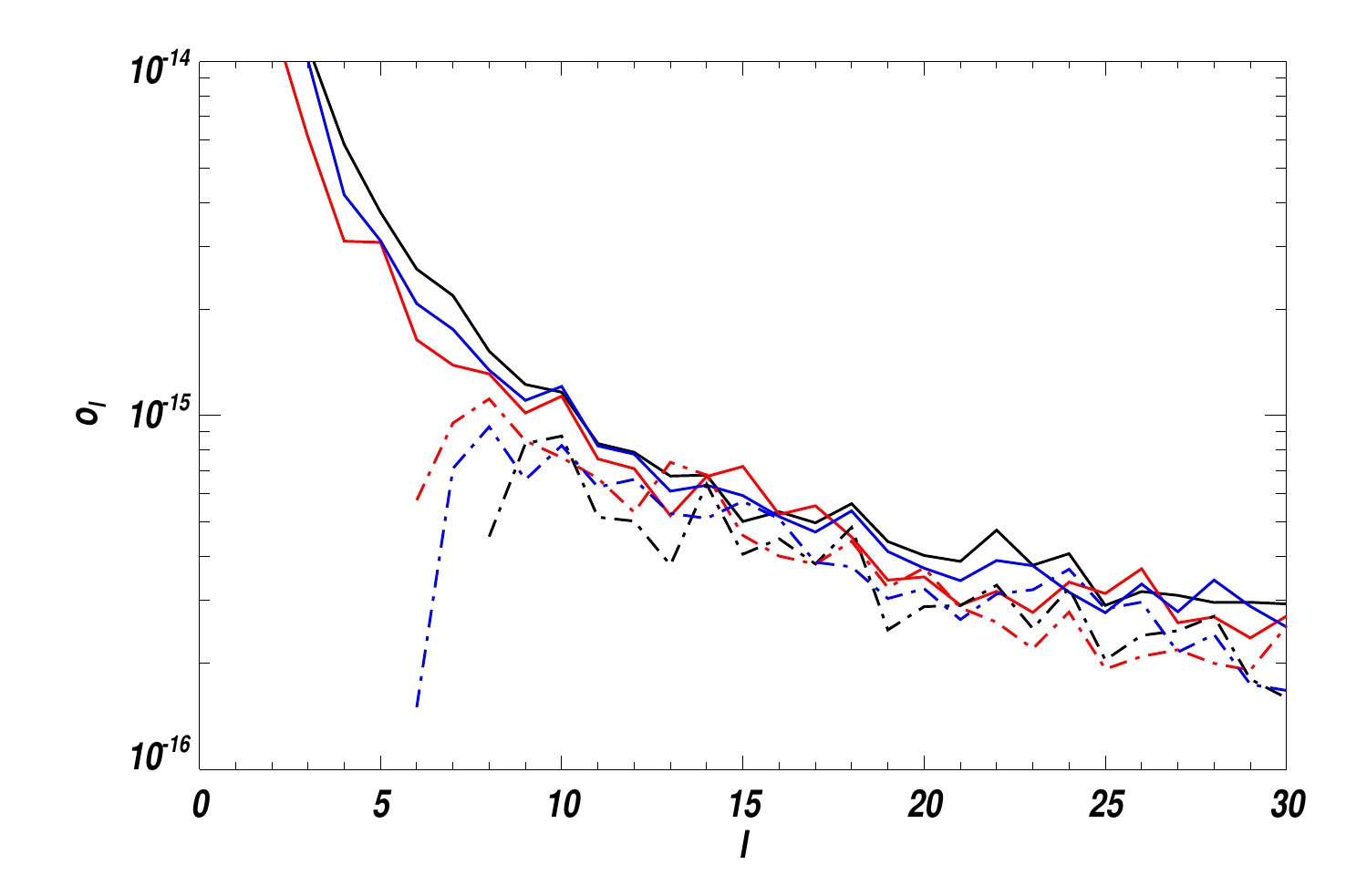}
\caption{The offset functions $o^{A\times B}_\ell$ of the oHL likelihood for the \hfi$\times$\hficq\ B-modes cross-spectra (solid) and the E-modes cross-spectra (dashed). The different colors refers to the different fiducial models used in the simulations: black is the early reionization scenario without tensor modes (model 1), red is the late reionization scenario with tensors (model 2) and blue is the \planck\ 2015 best fit with $r=0.1$.}
\label{Fig:offsetEEBB}
\end{figure}
Given that the offsets are needed to shift the $\cl$ distributions for each field T, E, B to avoid negative eigenvalues on the {\bf P} matrix, the offset functions depend on the shape of the $\cl$ distribution at each $\ell$. 
In particular, the offsets depend on the noise levels of the maps involved in the cross-spectra and on the mask used. In fact, the tails of the $\cl$ distributions at each $\ell$ are more negative when the noise is higher and when a larger mask is applied. 
The plot in \fig{Fig:offsetEE} shows how the offset functions change for different combinations of noise levels in the case of the six cross-spectra considered: from the highest of the WMAP$\times$\lfi\ in orange to the \hfic$\times$\hficq\ in green. 

Moreover, the offsets also depend on the fiducial model, as, in general, an higher signal-to-noise implies that the $\cl$ distributions have a smaller shift to negative values. \fig{Fig:offsetEEBB} shows the offset functions obtained from simulations generated with different fiducial models for the E-modes (dashed) and B-modes (solid) for the \hfic$\times$\hficq\ cross-spectra. The black lines refers to the early reionization scenario without tensor modes (model 1), the red lines to the late reionization scenario with tensors (model 2) and the blue lines to the \planck\ 2015 best fit with $r=0.1$. The shape of the offsets is consistent for the three different cases and, given the very different levels of signal considered, the dependence on the fiducial model is mild. As we will show in \sect{Sec:EEonly} and \sect{Sec:results_oHLcorr} the likelihood results on the cosmological parameters reconstruction are robust with respect to the choice of the fiducial model used to define the offset functions.

The $[M_f^{-1}]_{\ell \ell'}$ in \eq{eq:oHL} is the inverse of the cross-spectra $C^{A \times B}_\ell$-covariance matrix 
that can be estimated for a given theoretical fiducial model $C^{XY fid}_\ell$ through Monte Carlo simulations such that:
\begin{equation}
[M^{A\times B}_f]^{XY}_{\ell \ell '} =\langle \Big((C^{XY}_{\ell})_{sim}-C^{XY fid}_{\ell}\Big) \Big( (C^{XY}_{\ell '})_{sim}-C^{XY fid}_{\ell'}\Big) \rangle_{MC}, 
\label{eq:covmat}
\end{equation}
where $C^{XY}_{\ell} \equiv (C^{XY}_{\ell})^{A\times B}$, and $X,Y=\{T,E,B\}$. 

Since it will be useful in the following, we also report the equations of the modified oHL likelihood in the case of the single field approximation. In particular, we are interested in applying the method to the polarization EE-only cross-spectra $C_\ell^{EE} \equiv (C_\ell^{EE})^{A\times B}$ for which the oHL likelihood is defined by 
\begin{equation}
	-2\mbox{ln\lik}= \sum_{\ell \ell'} [OX_g]^{EE}_\ell [M_f^{EE}]^{-1}_{\ell \ell '} [OX_g]^{EE}_{\ell \ell'}
	\label{eq:oHLEE}
\end{equation}
where:
\begin{equation}
	[X_g]^{EE}_\ell \rightarrow [OX_g]^{EE}_\ell= \sqrt{ O(C_\ell^{EE fid}) } {\bf g} \Big[ \frac{ O(\hat{C}^{EE}_\ell)}{O(C^{EE mod}_{\ell })} \Big] \sqrt{ O(C_\ell^{EE fid}) },
	\label{eq:oHLtrasfEE}
\end{equation}
and:
\begin{equation}
	O(C_\ell^{EE})={(C^{EE}_\ell + o_\ell)}.
\end{equation}
$C_\ell^{EE fid}$, $\hat{C}^{EE}_\ell$ and $C^{EE mod}_{\ell }$ are, respectively, the spectra of the fiducial model, the data and the variable spectra for the likelihood sampling, and $o_\ell^{EE}$ is the effective offset.
Also, the covariance matrix to account for the multipole coupling in this case is defined by:
\begin{equation}
[M_f]_{\ell \ell '} =\langle \Big((C^{EE}_{\ell})^{A\times B}_{sim}-C^{EE fid}_{\ell}\Big) \Big( (C^{EE}_{\ell '})^{A\times B}_{sim}-C^{EE fid}_{\ell'}\Big) \rangle_{N_{sims}} 
\label{eq:covmatEE}
\end{equation}


\section{Single field results}\label{Sec:EEonly} 

\begin{figure*}
	\includegraphics[width=\linewidth]{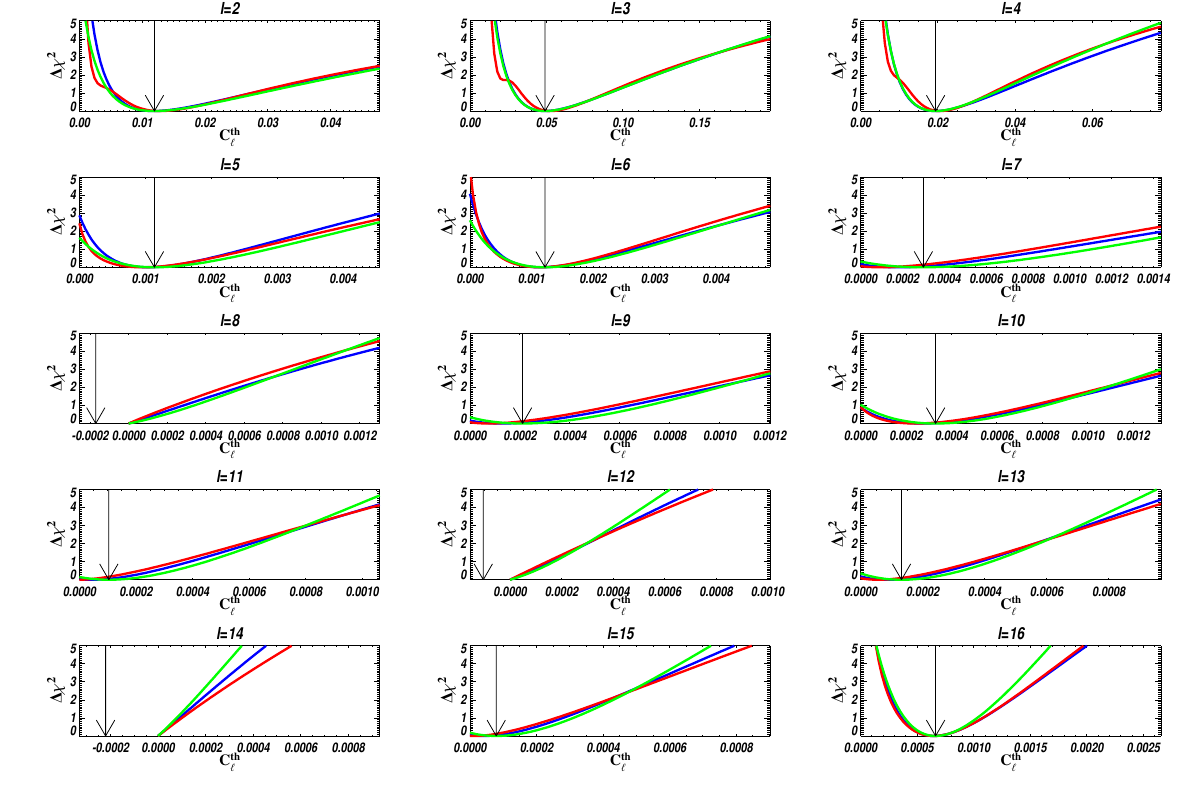}
	\caption{\label{fig:lklcomp} Comparisons of the three likelihoods methods developed for the low-correlations case ($\fsky=0.8$): red is for the Edgeworth expansion method, blue for the full-sky based method and green for the modified Hamimeche-Lewis approximation (oHL). The arrow represents where our random sample has fallen. Due to noise, it can be negative. The ordinate is $\Delta \chi^2=-2 \ln [\lik(\clth)/\lik_{max}]$.}
\end{figure*}

We first present the results in the case of the single field approximation. As single field we choose the E polarization and we build the EE cross-spectra likelihoods to constrain the $\tau$ parameter, since it is relevant for the analysis of present and future CMB data.
We construct the three different single field cross-spectra likelihoods derived from the formulas in \sect{Sec:cross-spectra}: the
general analytical parametrization derived from full-sky based approach, the parametrization based on the Edgeworth expansion
approximation to describe the cumulants of the cross-spectra distribution and the oHL single-field likelihood. 

In order to compare the three methods, we focus on the small sky cut case, where the cross-spectra simulations are generated by applying a mask with \fsky=0.8. The $\ell$-by-$\ell$ correlations are weak and the analytic approximations are reliable.
This comparison is useful not only as a validation test of the different methods but also to demonstrate that correlations can indeed be neglected in the parametric case.
To construct the single field oHL cross-spectra likelihood we use \refeq{oHLEE} where the $\ell$-$\ell$ correlations are encoded in the cross-spectra covariance matrix of \refeq{covmatEE}. For each of the six cross-spectra considered, the covariance matrix is computed from the Monte Carlo average of 10.000 E-modes simulations generated with a fiducial input cosmology corresponding to the \planck\ best-fit 2015 with $\tau=\tauvp$ and tensor modes with $r=0.1$. 
We estimate the offsets $o^{EE \, A\times B}_\ell$ from our reference simulations as described in \sect{Sec:theory_oHL} and \fig{Fig:offsetEE}.

Note that, as pointed out in \sect{subset:parametrization} and \sect{Sec:theory_oHL}, the parametrization used to define the analytical approximations and the definition of the offset functions of the oHL likelihood are both robust with respect to significative changes of the $\tau$ parameter in the fiducial model ($\Delta \tau_{fid} \simeq 0.03 >> \sigma_\tau$). 
However, in general, as it is the case for the HL likelihood analysis at smaller scales \citep{Hamimeche:2008ai}, it is a good choice to use a fiducial model close to the "true" model to compute the covariance matrix.

The likelihood sampling is done by computing the $C_\ell^{EE mod}$ with the CAMB code\footnote{http://camb.info/} 
varying $\tau$ in the range $[0.01,0.15]$ with a step $\Delta \tau=0.001$, fixing the other parameters to their \planck\ 2015 best-fit values and rescaling $A_s e^{-2 \tau}$. The degeneracy between $\tau$ and the scalar amplitude parameter $A_s$ is in fact broken by fixing accordingly the amplitude of the first peak of
the TT spectrum $A_{TT}=A_s e^{-2 \tau}$ at $\ell = 200$. More general results based on joint constraints of the $\tau$ and $A_s$ are presented in \sect{sub:tauas}. 

%
To compare the three likelihoods, we choose events (\ie one \cl\ vector sample) at random from the set of \hfic$\times$\hficq\ simulations
 and construct for each $\ell$ independently the
marginal likelihoods with the three different methods, setting each time all \clth\ values other than this multipole to their true values.
Fig. \ref{fig:lklcomp} displays a typical case. Here are some
comments that we derive from the observation of many samples: even-though the sample \cl\ may get negative values, due to noise
  and low signal, the likelihood of any negative true power value
  is unphysical and is equal to 0. This case does not happen in
  practice since in cosmological parameter estimation the Boltzmann
  code always propose positive spectra. The Edgeworth-based method shows some oscillation for the very first
  multipoles, generally for $\ell=2,3$.  
  This is due to the very steep raising of the
  distributions at the very beginning
  (see Fig \ref{fig:distrib5_anal}) which leads to some small
  negative \q{ringing} effect in the truncated expansion. 
  The method introduced in \sect{subsec:results_edge} mitigates the
  effect but does not completely cure for it and regarding this aspect the full-sky based method and the oHL likelihood gives a better approximation. 
 Overall \fig{fig:lklcomp} shows the excellent agreement among the three likelihood methods in recovering the minimum $\Delta \chi^2=-2 \ln [\lik(\clth)/\lik_{max}]$ for each multipoles with comparable accuracy.

As a further validation test, we check the bias of the likelihood against our set of 10.000 Monte-Carlo simulations. For each simulation, we derive the distribution of the maximum likelihood of $\tau$ for $\ell<20$. For the full sky based likelihood and for the Edgeworth expansion likelihood we remove multipoles 2 and 3 -which do not carry much information due to the cosmic variance level- since their p.d.f parametrization is less accurate, as shown in \sect{subset:parametrization}. For oHL we consider $\ell=[2,20]$.

\begin{figure}
	\centering
	\includegraphics[width=0.92\linewidth]{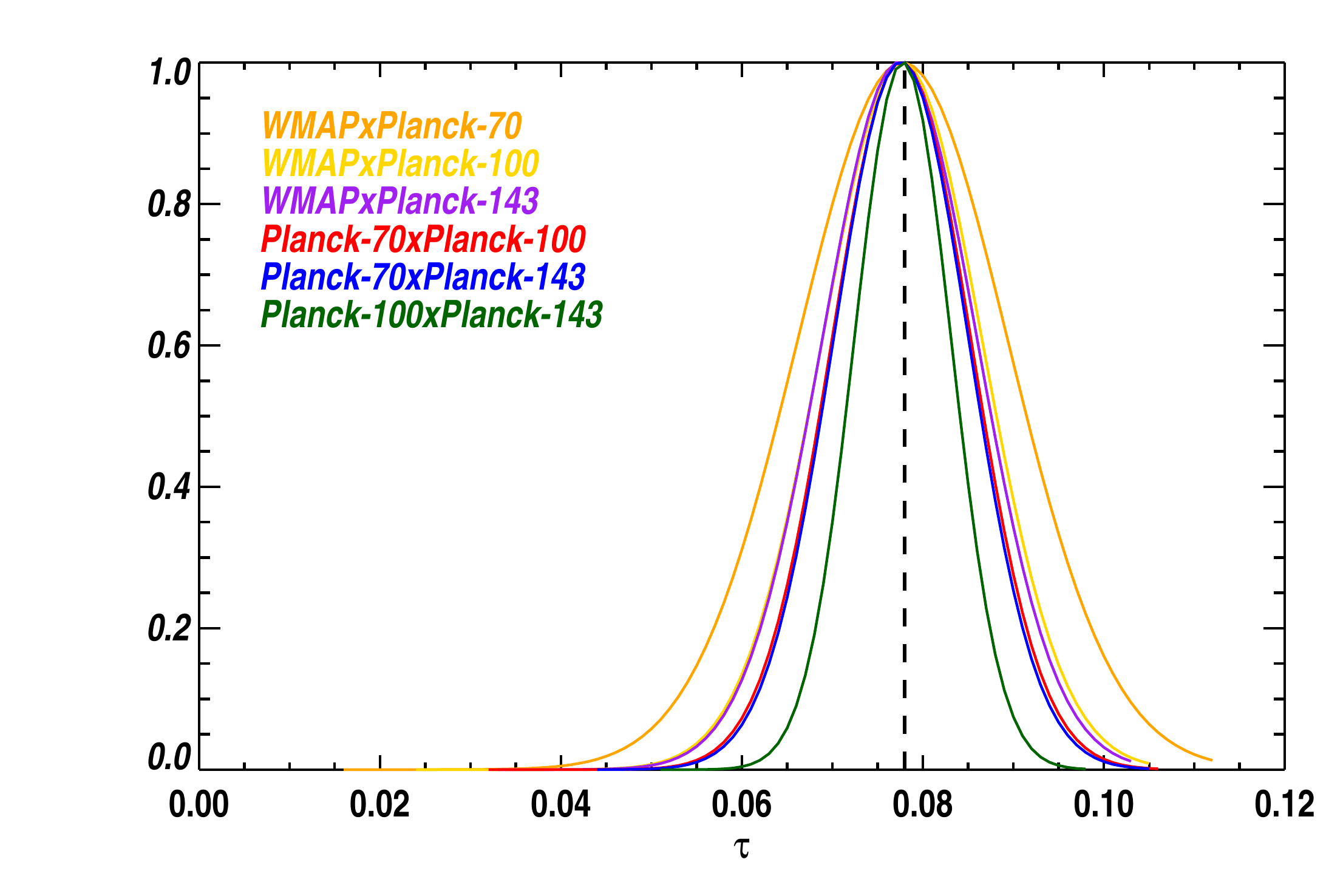} \\
	\includegraphics[width=0.92\linewidth]{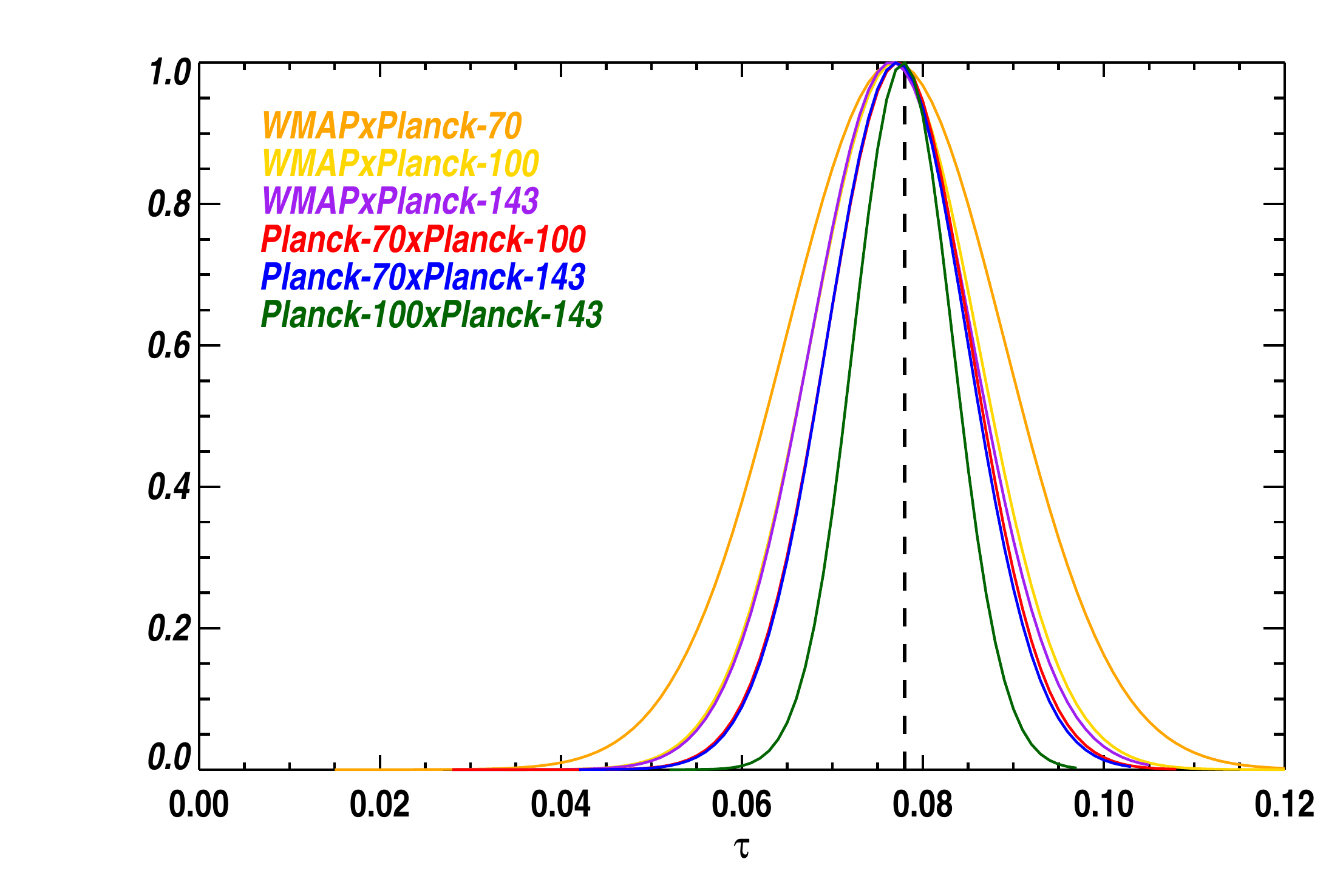} \\
	\includegraphics[width=0.92\linewidth]{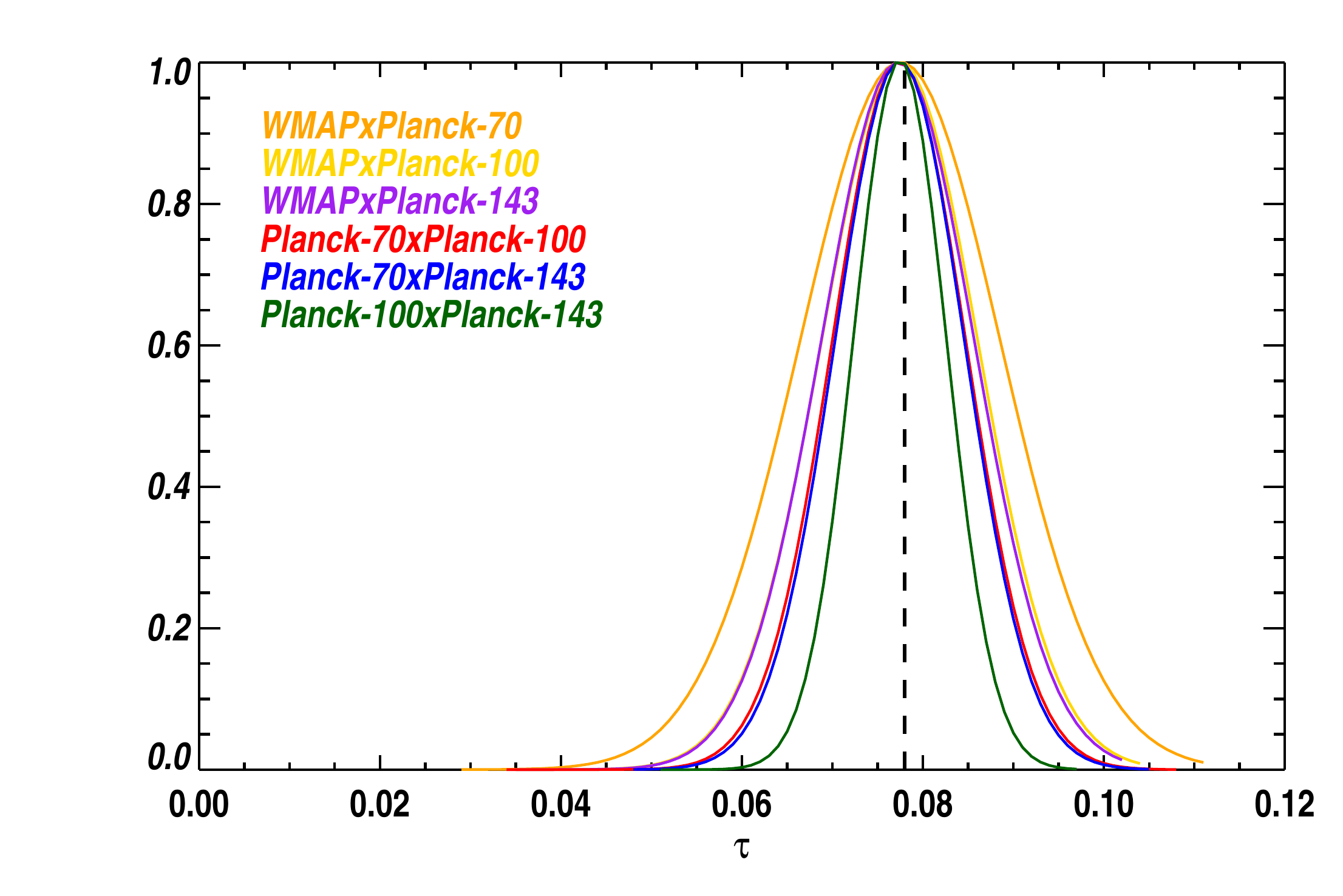}
	\caption{Distribution of the maximum probability for the analytic full-sky based likelihood ({\it top}), for the analytical parametrization based on the Edgeworth expansion ({\it middle}) and for the oHL likelihood ({\it bottom}) on each E-modes cross-spectra on 80\% of the sky}.
	\label{fig:dist_tau_lik}
\end{figure}

Figure~\ref{fig:dist_tau_lik} shows the distribution of the maximum probability over the Monte Carlo simulations for the full sky based likelihood, the Edgeworth expansion likelihood and the oHL likelihood. All three approximations recover the input value $\tau_{fid}=0.078$ used to generate our reference simulations, showing that the three likelihoods are unbiased.
In \tab{tab:lik_errors} are reported the best fit values and the error bars on the estimation of the $\tau$ parameter. The error bars are computed as the standard deviation of the maximum probability $\tilde\tau$ for the three likelihoods. Since the oHL likelihood accounts for the $\ell$-by-$\ell$ correlations while the full-sky based likelihood and the Edgeworth expansion approximation do not, the fact that 
the three methods give compatible results in terms of error bars confirm that the level of multipole correlations for a small sky cut is low and does not have an impact in the reconstruction of the $\tau$ parameter. Note however that both the analytical approximations are slightly sub-optimal with respect to the oHL likelihood, by a factor of $\simeq 4\%$ for the full-sky based likelihood and by a factor of $\simeq 7\%$ for the Edgeworth expansion likelihood. These results hold in general for all the cross-spectra considered.
\begin{table*}
	\caption{ Comparison of the best fit estimation of the reionization optical depth $\tilde\tau$ and error bars $\sigma_{\tilde\tau}$ for the three likelihood methods from simulations (see also \fig{fig:dist_tau_lik}). The errors are computed as the standard deviation of the maximum probability $\tilde\tau$ over a set of 2000 simulations. The input value used in the simulations is $\tau_{fid}=0.078$ and $\fsky=0.8$.}
	\center
	\begin{tabular}{cccc} 
	\hline
	Cross-spectra & $(\tilde\tau \pm \sigma_{\tilde\tau})^{fullsky}$ & $(\tilde\tau \pm \sigma_{\tilde\tau})^{edgeworth}$ & $(\tilde\tau \pm \sigma_{\tilde\tau})^{oHL}$ \\
	\hline
WMAP$\times$\lfi & $0.0777\pm 0.0116 $& $0.0768\pm0.0121 $&    $0.0774\pm0.0110$ \\
WMAP$\times$\hfic & $0.0777\pm0.0088$ & $0.0768\pm0.0092 $ & $0.0773\pm 0.0086$  \\
WMAP$\times$\hficq & $.00774\pm0.0086 $& $0.0765\pm0.0089$  & $0.0772\pm 0.0084 $ \\
\lfi$\times$\hfic & $ 0.0776\pm0.0077$ & $0.0773\pm0.0079$ &  $0.0774\pm 0.0074$  \\
\lfi$\times$\hficq&  $0.0775\pm0.0075 $ & $0.0771\pm0.0078 $ & $0.0774\pm0.0071$  \\
\hfic$\times$\hficq & $0.0777\pm0.0054$ & $0.0778\pm0.0055$ &  $0.0781\pm0.0051$  \\
	\hline
	\end{tabular}
	\label{tab:lik_errors}
\end{table*}
\begin{figure}
	\centering
	\includegraphics[width=7.5cm,height=4.5cm]{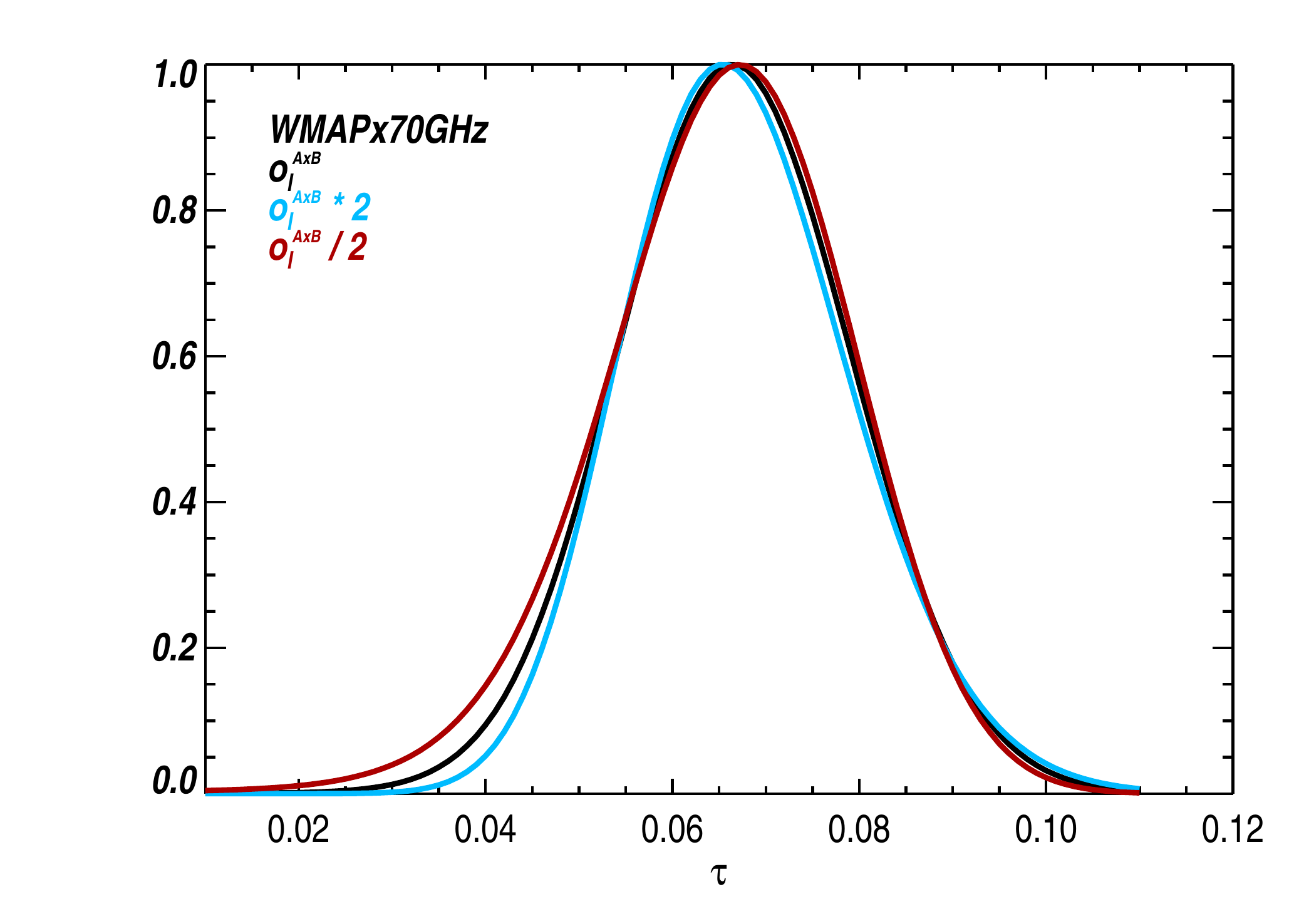}\\
	\includegraphics[width=7.5cm,height=4.5cm]{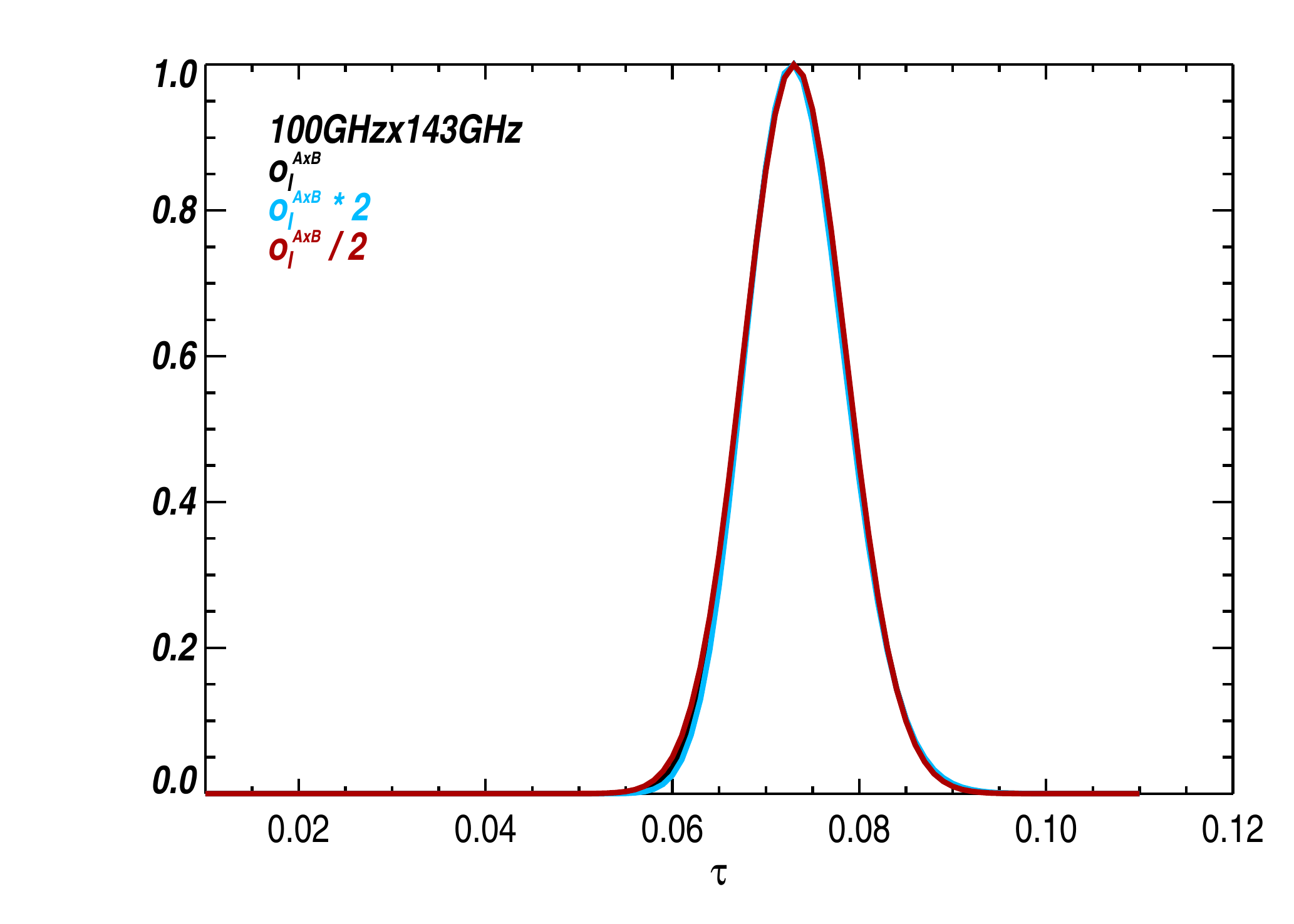}
	\caption{The plots show the effect of changing the offsets amplitude by a factor of two on the posterior distribution of the $\tau$ parameter. 
	The top panel refers to the \wmap $\times$\lfi\ cross-spectra, while the bottom panel to the \hfic$\times$\hficq\ cross-spectra,  
	}
	\label{fig:posterior_offset}
\end{figure}

Finally, it is useful to assess the stability of the results obtained with the oHL likelihood with respect to choice of the offset term.
Indeed, changing the offsets both could bias the peak of the posterior distribution and change its width. 
As described in \refeq{oHLtrasfEE}, the offset ensure the H\&L transformation to be definite and too small offsets may leak to undefined likelihood. On the opposite, a overestimation of the offset value has limited effect on the peak distribution.
Figure~\ref{fig:posterior_offset} shows that the impact of a factor of two in the estimation of the offsets amplitude is negligible on the posterior distribution of the $\tau$ parameter. The figure illustrates two representative cases of $\tau$ posteriors obtained with the highest and lowest noise configuration from our simulations. 
Note that a change in the offset of this type could arise if the fiducial model used to derive the offsets is very different from the best fit model, as illustrated in \fig{Fig:offsetEE}. The fact that this change has practically no effects on the posterior distributions demonstrates that the definition of the oHL likelihood is robust with respect to the offset reconstruction.
Also, the effect on the width of a change in the offset is very weak, meaning that the offsets, as expected, do not affect the estimation of the error bars. The same results hold true for all the cross-spectra considered.
In general, our offset terms are well defined and the oHL results are robust with respect to the offset choice. 

For a smaller sky coverage, the three likelihoods remain unbiased even if the $\ell$-by-$\ell$ correlations get larger. 
In this case, the parametric likelihoods are less optimal: their variance increase by about 10\% compared to the oHL likelihood for the 50\% mask.


\section{Results for correlated fields}\label{Sec:results_oHLcorr}

\begin{figure}
	\centering
	\includegraphics[width=7.5cm,height=5.5cm]{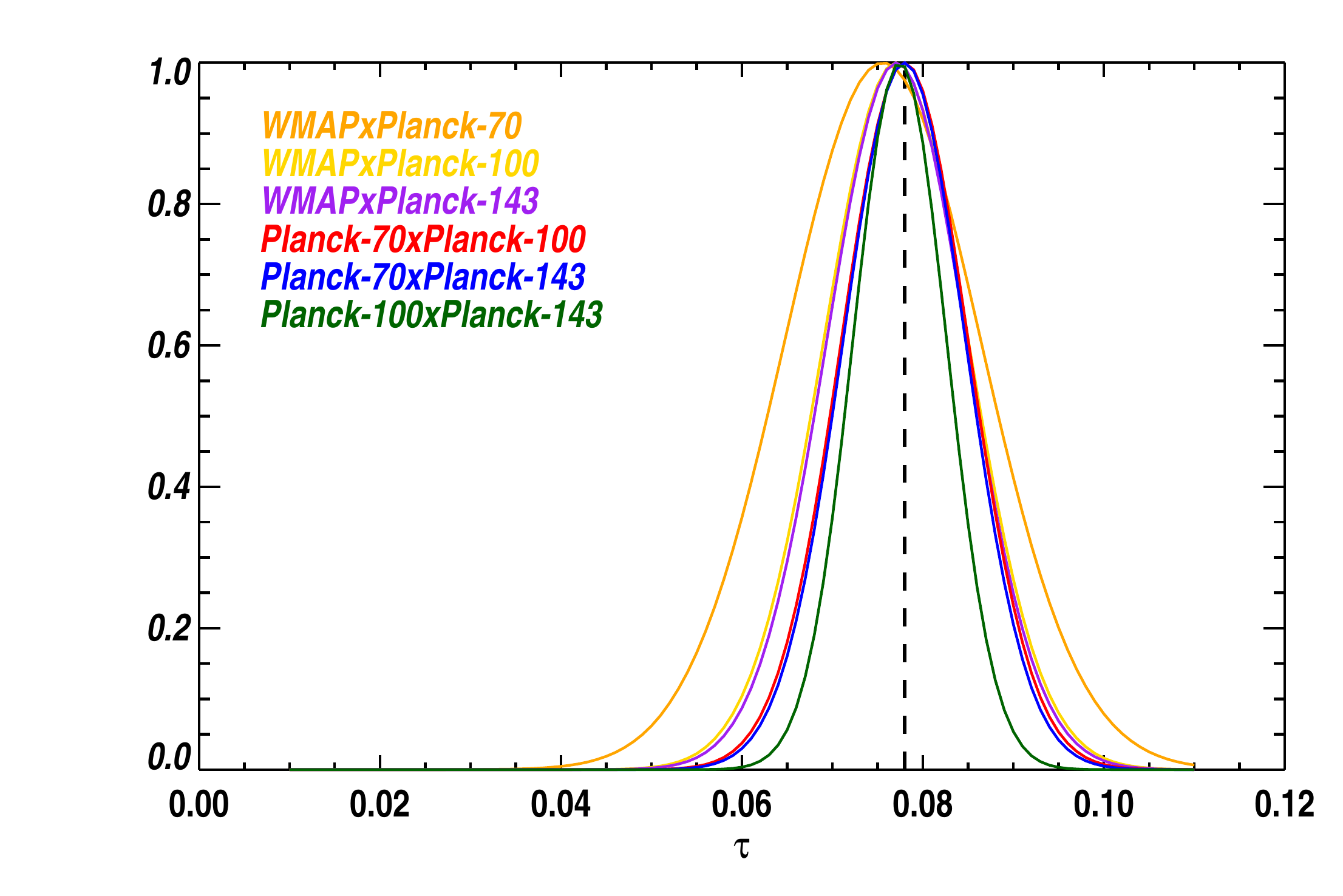}\\
	\includegraphics[width=7.5cm,height=5.5cm]{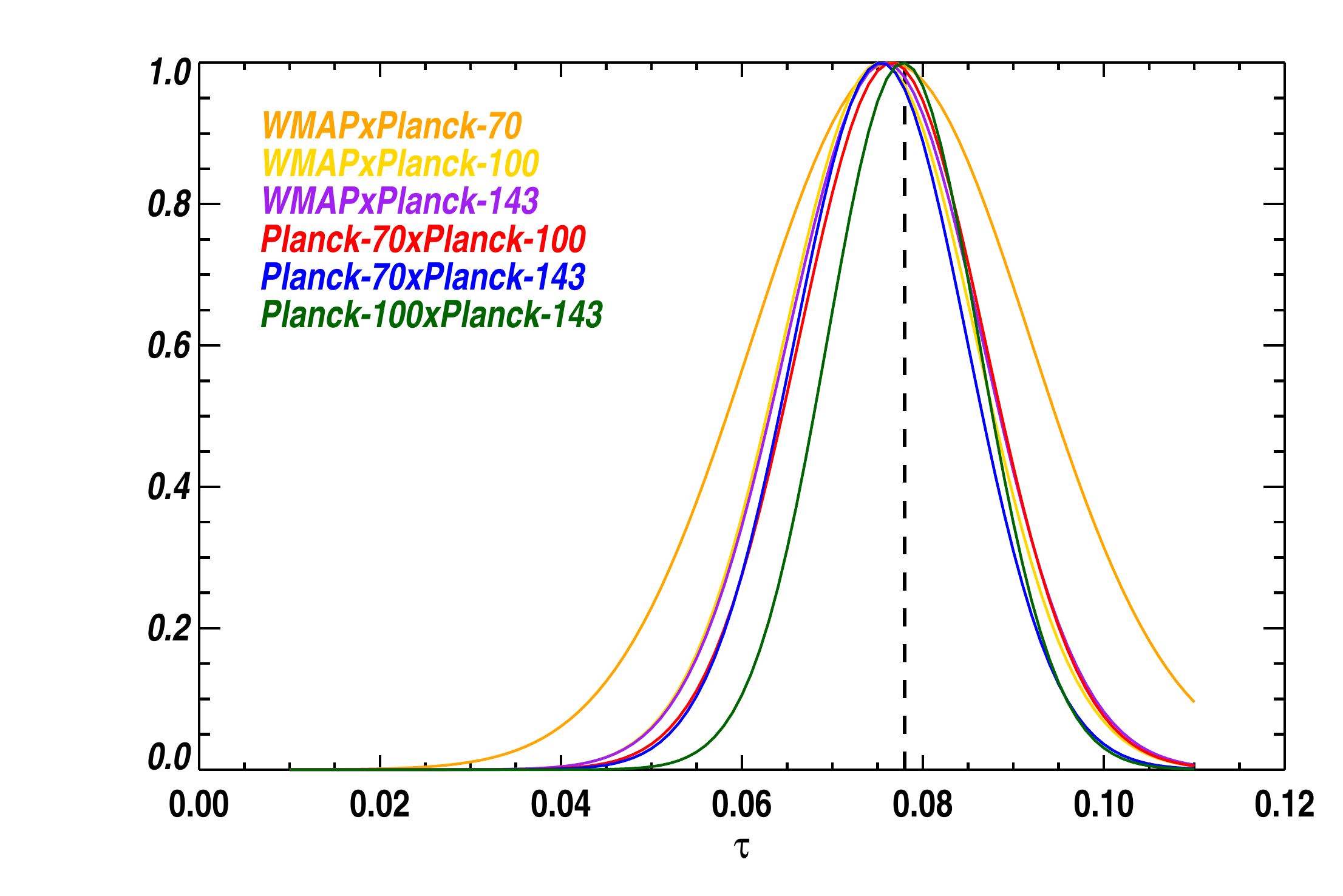}
	\caption{{\it Validation of the oHL multi-fields likelihood.} The plots show that the oHL likelihood computed combining the T, E and B fields and accounting for both multipole and fields correlations gives unbiased results on the estimation of the optical depth to reionization parameter $\tau$. The top panel shows the $\tau$ posterior for the six different cross-spectra when 20\% of the sky is masked ($f_{sky}=0.8$), while the bottom panel shows the results for a bigger mask with $f_{sky}=0.5$. The dashed line refers to the input value $\tau_{fid}=\tauvp$ used in the simulations.}
	\label{fig:simsOHLallcorr}
\end{figure}

This section is dedicated to the results obtained with the full temperature and polarization oHL likelihood (\refeq{oHL}).
One of the main advantages of the oHL method relies in fact on the possibility to include in the analysis both the correlations between the [T,E,B] fields and the multipole correlations.
Since the simulations used for each cross-spectrum are built with realistic noise levels as described in \sect{Sec:pcl}, the forecasted estimates on the $\tau$, r and $A_s$ parameters from the low-$\ell$ analysis presented here are realistic for current CMB experiments.

We build the \{TT, EE, BB, TE, TB, EB\} oHL likelihood for the six different cross-spectra: \wmap$\times$\lfi, \wmap$\times$\hfic, \wmap$\times$\hficq, \lfi$\times$\hfic, \lfi$\times$\hficq, \hfic$\times$\hficq. For each cross-spectrum we construct the full [T,E,B] covariance matrix of \eq{eq:covmat} by computing the Monte Carlo average of 10.000 simulations generated with a fiducial input cosmology corresponding to our baseline \planck\ 2015 best fit with $\tau=\tauvp$ and $r=0.1$. The offsets functions are derived from the same simulations as described in \sect{Sec:theory_oHL}. 
For each cross-spectra we therefore add the offsets $o^{TT}_\ell$, $o^{EE}_\ell$, $o_\ell^{BB}$ to the diagonal elements of the $C_\ell$ matrix as defined in \eq{eq:oHLgeneral}.

\subsection{Constraints on $\tau$} 
Firstly, we study the impact of including the T,E,B cross-spectra and their correlations on the estimation of the optical depth to reionization $\tau$, compared to the single-field EE analysis described in the previous section \sect{Sec:EEonly}. The sky fraction is $f_{sky}=0.8$ and the multipole range used is $\ell=[2,20]$.
As shown in \fig{fig:simsOHLallcorr} and Table \ref{t:oHLallcorr}, the combined analysis gives unbiased results on the estimation of $\tau$. 
As expected, adding the temperature and the tensor modes and all the possible correlations gives results very close to the single-field EE analysis since the relevant physical information related to $\tau$ is essentially encoded in the EE-spectra. However, the full temperature and polarization analysis leads to a slight improvement in the estimation of the $\tau$ error bars, which is of about a few percent for the \hfic$\times$\hficq\ analysis. 

\begin{figure}
	\centering
	\includegraphics[width=7.5cm,height=5.5cm]{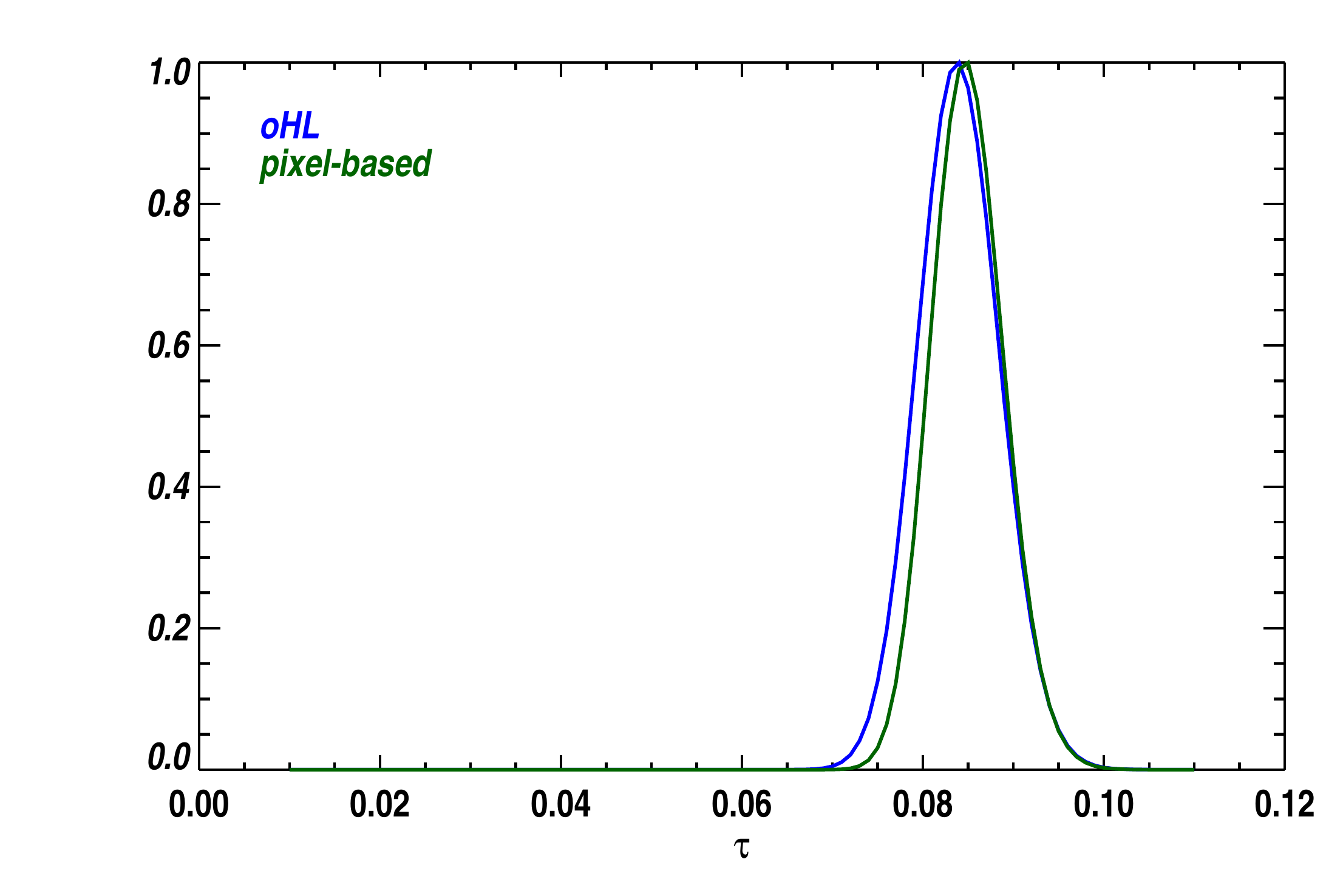}
	\caption{Comparison of the posterior distributions of the $\tau$ parameter obtained with the full temperature and polarization oHL likelihood (blue) and with the pixel-based likelihood (green). The plot shows a typical example from the \hfi\ simulation set.}
	\label{fig:pixelVSoH}
\end{figure}

We compare the $\tau$ posterior distribution of the oHL likelihood to the one from the pixel-based likelihood. 
We implement a pixel-based likelihood by using a combination of the maps at \hfic\ and \hficq\ from the same simulation set that we used to generate the \hfi\ cross-spectra. Both methods are therefore based on simulations with the same noise characterization. A typical case is given in \fig{fig:pixelVSoH}.
As expected, the oHL likelihood approximation is slightly sub-optimal with respect to the pixel-based likelihood which is not an approximation and is build to be statistically optimal. Note however that the error bars obtained with the oHL likelihood are comparable with the optimal estimate obtained by using the pixel-based approach at better than 15\%.

Finally, we use the combined oHL likelihood to test the results with a different sky cut. 
We consider a severe cut at 50\% ($f_{sky}=0.5$). This is a more complicate case to deal with since the $\ell$-by-$\ell$ correlations are stronger. %
Also, the shape of the distributions of the $\cl$ estimators at each $\ell$ is affected by the smaller sky coverage, leading to more negative tails. We generate the offset functions for each cross-spectra as described in \sect{Sec:theory_oHL}, using our reference simulations masked at 50\%.
The results are summarized in table~\ref{t:oHLallcorr} and in the bottom panel of \fig{fig:simsOHLallcorr} that shows the $\tau$ posteriors for each of the six cross-spectra. Even in this more complex case, the oHL likelihood analysis is unbiased.
As expected, since we are considering a smaller sky fraction and non-negligible multipole correlations, we recover bigger error bars with respect to the $\fsky=0.8$ analysis, with a degradation of $\simeq 30\%$ for the \hfic$\times$\hficq.

\begin{table}
\caption{
Results on the estimation of the $\tau$ parameter with the full temperature and polarization oHL likelihood. 
The fiducial model used in the simulation is the \planck\ 2015 $\Lambda$CDM best fit with $\tau_{fid}=0.078$. 
The table shows the comparison between the $\tau$ estimates (best fit $\tilde{\tau}$ and error bars $\sigma_{\tilde\tau}$) obtained with two set of simulations with different sky cuts: the small mask with with $f_{sky}=0.8$ and a bigger mask with $f_{sky}=0.5$.}
\begin{center}
\begin{tabular}{ccc} 
\hline
Cross-spectra & $\tilde\tau \pm \sigma_{\tilde\tau}$ ($f_{sky}=0.8$)   & $\tilde\tau \pm \sigma_{\tilde\tau}$ ($f_{sky}=0.5$)  \\
\hline
WMAP$\times$\lfi&    $0.0750 \pm 0.0108$ &$0.0761\pm 0.0203$\\
WMAP$\times$\hfic&   $0.0769 \pm 0.0075$ & $0.0764\pm 0.0121$\\
WMAP$\times$\hficq  & $0.0769 \pm 0.0079$ & $0.0770\pm 0.0116$\\
\lfi$\times$\hfic    & $0.0783\pm 0.0069$ & $0.0776 \pm 0.0105$\\
\lfi$\times$\hficq   &  $0.0784 \pm 0.0065$  &$0.0763\pm 0.0101$\\
\hfic$\times$\hficq  &$0.0780 \pm 0.0049$ & $0.0788\pm 0.0069$\\
\hline
\end{tabular}
\label{t:oHLallcorr}
\end{center}
\end{table}

\subsection{Joint estimation of $\tau$, r and $A_s$}\label{sub:joint2D}
Using the full combined analysis, we can construct multi-dimensional constraints on parameters. In particular, we focus on the correlations between the optical depth and the amplitude of the scalar fluctuations $A_s$ and between the optical depth and the tensor-to-scalar ratio $r$ which are relevant for the future analysis of CMB data at large angular scales from e.g. \planck.
In both cases we perform the full analysis using the \hfic$\times$\hficq\ spectra which corresponds to the lowest noise frequency combination and it can be used to make realistic forecasts for current and future CMB experiments. We consider a sky cut with $f_{sky}=0.8$.

\begin{figure*}
\centering
\includegraphics[width=8.8cm,height=6.5cm]{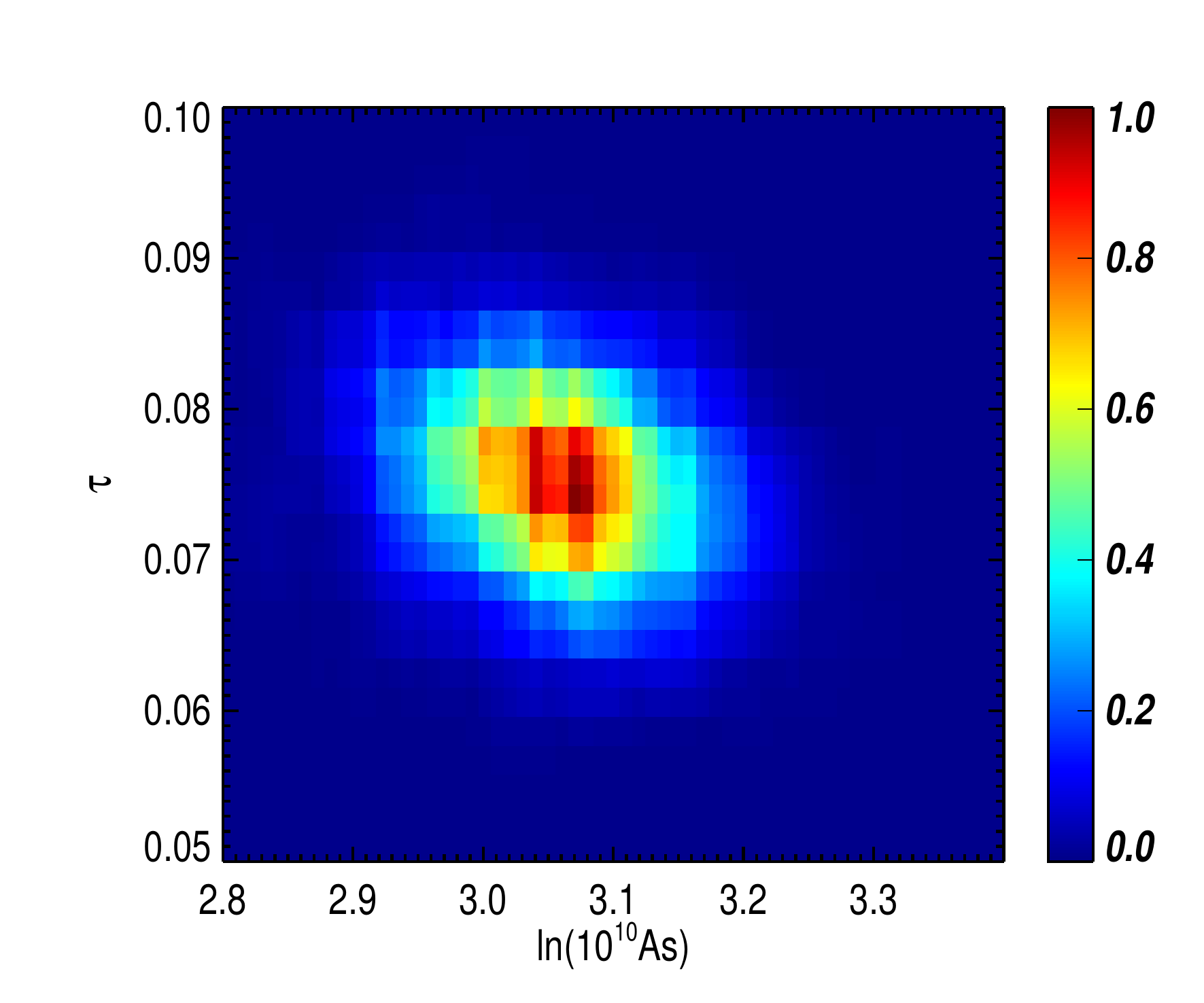}
\includegraphics[width=8.8cm,height=6.5cm]{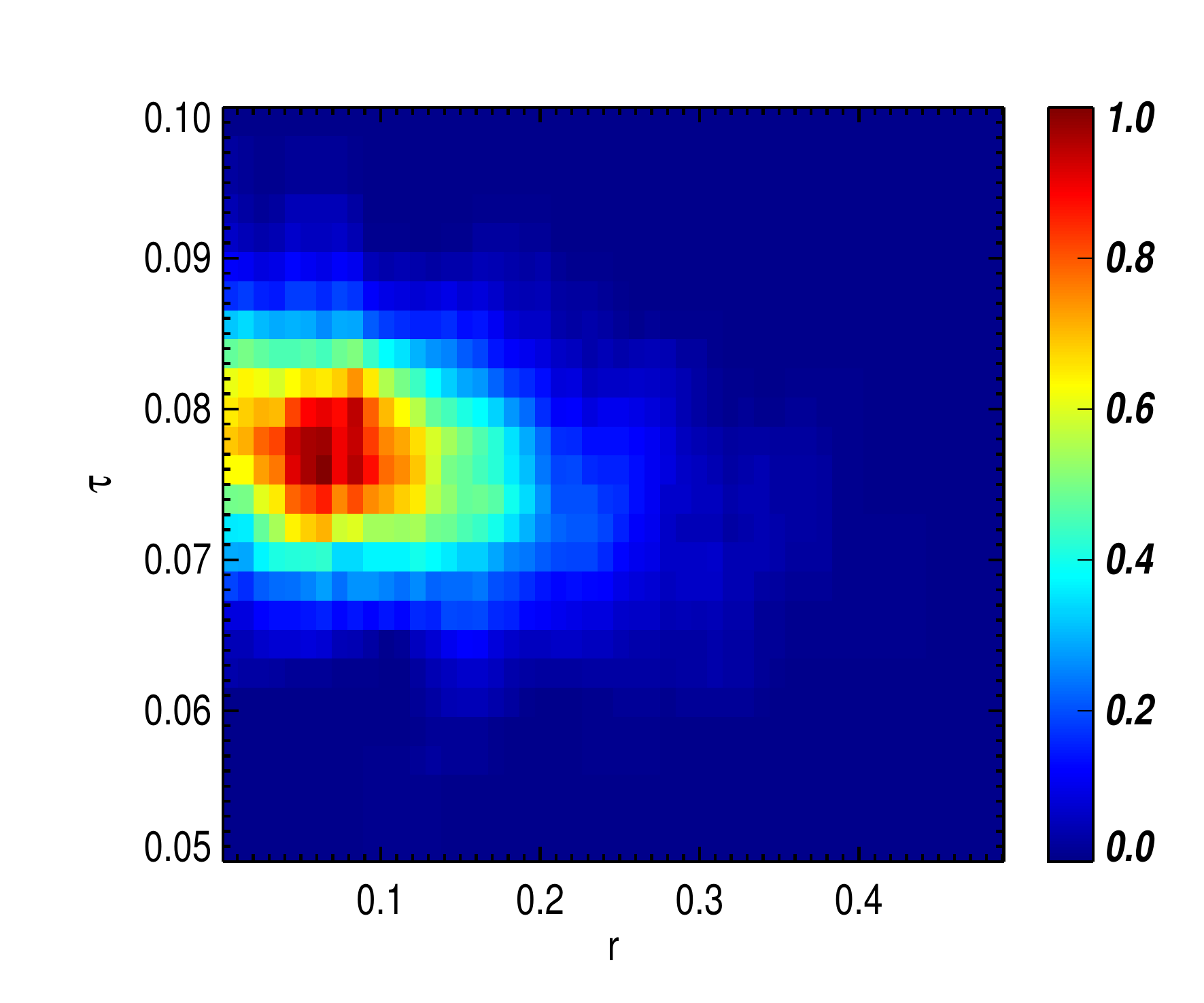}
\caption{{\it 2D-distribution of the maximum likelihood for $\tau$-$r$ (right panel) and $\tau$-$A_s$ (left panel)}. The plots show the joint constraints obtained with the full temperature and polarization oHL likelihood on 2000 simulations of the \hfic$\times$\hficq\ cross-spectra. The fiducial input parameters used in the simulations are: $\tau_{fid}=0.078$, $r_{fid}=0.1$ and $[ln(10^{10} A_s)]_{fid}=3.09$.}
\label{fig:2Dfit}
\end{figure*}

\subsubsection{Joint estimation of $\tau$ and $A_s$}\label{sub:tauas}
Using the temperature power spectrum only, $A_s$ and $\tau$ are strongly degenerated. Indeed, the amplitude of the first acoustic peak of the CMB temperature power spectrum directly measures $A_{TT}=A_s e^{-2\tau}$. 
Using polarization data at large angular scale helps breaking this degeneracy.
So far we fixed the degeneracy direction by rescaling the temperature spectrum, fixing $A_{TT}$, accordingly to the variation of $\tau$ in the likelihood sampling. Here we let $A_s$ free to vary.
The results from the simulations, using the \planck-100$\times$\planck-143 full oHL likelihood are summarized in the left panel of \fig{fig:2Dfit}. The plot shows the 2D histogram of the best fit values for the whole set of simulations in the $\tau$-$A_s$ projection. The full oHL likelihood correctly recovers the inputs values for $\tau$ and $A_s$ as well as error bars compatible with the MC dispersion.

\subsubsection{Joint estimation of $\tau$ and r}\label{sub:taur}
The CMB power spectra at large angular scales, in particular the E and B polarization modes, are affected by how the reionization process proceeded and lasted.
Thus, as shown in \fig{fig:models}, the power at large scales (low-$\ell$) in the B-modes spectrum is directly related to the reionization optical depth. Indeed, the amplitude of the B-modes spectrum reionization bump scales with $\tau^2$: $C^{BB}_{\ell<20}(\tau)\propto \tau^2 C^{BB}_{\ell<20}$. As the amplitude of the B-modes spectrum of course also depends on the amount of the primordial tensor perturbations, there is a degeneracy between the $\tau$ and $r$. It is therefore interesting to derive joint estimates of these parameters.

We compute the joint $\tau$-$r$ constraints with the full oHL likelihood on the set of 2000 simulations of the \hfic$\times$\hficq\ cross-spectra with an input cosmology corresponding to the \planck\ 2015 best fit for the base $\Lambda$CDM parameters with $\tau=\tauvp$ and a  tensor-to-scalar ratio of $r=0.1$. The multipole range used is, as usual, $\ell=[2,20]$.
The results of the oHL likelihood sampling on simulations are summarized in the right panel of \fig{fig:2Dfit}. The plot shows the posterior in the $\tau$-$r$ plane from the oHL and from which we can see that the oHL likelihood correctly recovers the parameters $\tau$ and $r$ compatible with the input values used in the simulations.
As regarding the error bars, the forecasted $1\sigma$ error for $\tau$ in the case of the highest resolution channels of a \planck-like experiment is $\sigma_\tau^{100\times143}=0.0051$.
For the tensor-to-scalar ratio in the multipole range considered, we find $\sigma_r^{100\times143}=0.09$.
Note that in our analysis, 
we consider a correlated noise model. This noise characterization, which is more realistic with respect to a simpler white noise modeling, implies a rising of the noise level at low multipoles due to the $1/f$ noise correlations (see \fig{fig:models}). Therefore, in particular in the case of a low signal scenario, the correlated noise at large scales can eventually dominate over the cosmic variance inducing a worsening of the constraining power proportional on how steep is the rising of the correlated noise at low multipoles.


\section{Conclusions}\label{Sec:conclusions}
In this paper we presented a new approach for the analysis of the CMB polarization data at large angular scales based on cross-correlation in spectra domain. Using cross-spectra with respect to the auto-spectra and, in general, to the pixel based approach used so far in the CMB analysis at large angular scales has many advantages, in particular in the case of a realistic CMB experiment that accounts for anisotropic noise and a sky cut needed to minimize the foreground contamination. In fact, by using cross-frequency/cross-dataset CMB spectra, the noise biases and the systematics specific to a given frequency/dataset are removed. Also, the possible foreground residuals can be minimized and the information encoded in different frequencies/datasets can be combined efficiently.

The cross-spectra estimators are non-Gaussian at low multipoles especially when applied on cut-sky. We generalized the approximation made in \cite{Hamimeche:2008ai} to accommodate for cross-spectra at very low multipoles.
This likelihood (oHL) can easily handle the correlation between CMB modes ($TT$, $EE$, $BB$, $TE$ as well as $TB$ and $EB$) and between multipoles and gives error bars less than 15\% larger than the optimal pixel-based method. The oHL likelihood shares the same robustness with respect to the choice of the fiducial model as the H\&L approximation (see discussion in \cite{Hamimeche:2008ai}). We compared the oHL likelihood to the analytical parametrization of the estimator distribution which can be used as a quick likelihood solution in the case of a single field analysis with small sky cuts so that correlations can be safely neglected. 

We generated different sets of simulations that we used to construct and validate the likelihoods, proving that all the methods are unbiased and can accurately constrain the optical depth to reionization parameter $\tau$. Also, we showed that the oHL likelihood gives accurate constraints of the joint estimation of the $\tau$ parameter, the tensor-to-scalar ratio parameter $r$ and the amplitude of the primordial scalar perturbations $A_s$. Our simulations account for anisotropic correlated noise, beam, mask with the characteristic of a realistic CMB experiment as \wmap\ and \planck. In order to validate our likelihoods for different noise levels, we generated simulations for cross-frequency spectra with different resolution, from the lowest, \wmap x\lfi, to highest, i.e. \hfic x\hficq. 

Optimal foreground cleaning is beyond the scope of this paper but foreground residuals, in particular synchrotron and dust, must be quantified in a realistic CMB analysis. In this paper we work with cleaned CMB maps but we account and propagate the uncertainties related to the foregrounds removal by using in our simulations realistic estimates derived from public data.
The correlated noise term that we include in the simulations in fact is drown from real data and can be taken as a good proxy for a realistic combination of noise, systematics and foregrounds residuals, in particular at low multipoles. 
 
The cross-spectra likelihood approach presented in this paper is a powerful and efficient tool for the analysis of the CMB data at large angular scales. It allows to minimize the impact of the experimental residual systematics (from both instruments and foreground contamination) while providing nearly-optimal constraints on the estimation of the $\tau$, $r$ and $A_s$ cosmological parameters. 

\section*{Acknowledgements}

We acknowledge Antony Lewis for useful discussions on the cross-spectra statistics and likelihood. We acknowledge Olivier Perdereau, Marta Spinelli and Sophie Henrot-Versille for useful comments. 

\bibliographystyle{Bib/mn2e} %
\bibliography{Bib/lowlpol}

\newpage

\appendix

\section{Cross-spectra distribution on the full sky}
\label{app:cross_distrib}

We consider the product of a pair of correlated central gaussian random variables:
\begin{equation}
  x=z_a \times z_b \qquad (z_a,z_b) \sim {\mathcal N_2}(\vec{u};\vec{0},\mathbf{V})
\end{equation}
where $\vec{u}$ is a generic vector and the covariance matrix is written in the standard form:
 \begin{equation}
\mathbf{V}=
\begin{pmatrix}
  \sa^2 & \rho \sa \sb\\ 
  \rho \sa \sb & \sb^2
\end{pmatrix}.
\end{equation}

Standard probability rules allow to compute its \pdf \citep{Grishchuk:1996}:
\begin{equation}
\label{eq:prod1}
  f_x(x)=\dfrac{1}{\pi\sa\sb\sqrt{1-\rho^2}}e^{\dfrac{\rho x}{(1-\rho^2)\sa\sb}}K_0\left(\dfrac{\abs{x}}{(1-\rho^2)\sa\sb}\right)
\end{equation}
whose characteristic function (Fourier transform is):
\begin{equation}
\label{eq:char1}
  \phi_x(t)=\E{e^{i x t}}=\dfrac{1}{\sqrt{1-2i \rho \sa \sb t + (1-\rho^2)\sa^2\sb^2 t^2}}.
\end{equation}

The sum of $N$ such independent variables $X=\sum_{i=1}^N x_i$ has therefore the characteristic function:
\begin{align}
  \phi_X(t)=&\left[1-2i \rho \sa \sb t + (1-\rho^2)\sa^2\sb^2 t^2) \right]^{-N/2} \\
  & \left[ (1-\rho^2)\sa^2\sb^2(t-\dfrac{i}{(1-\rho)\sa\sb})(t+\dfrac{i}{(1+\rho)\sa\sb})\right]^{-N/2}. \nonumber
\end{align}

To obtain the $X$ \pdf we inverse-Fourier it:
\begin{align}
  f_X(x)=&\dfrac{1}{2\pi} \int_{-\infty}^{+\infty} \phi_X(t) e^{-i x t} dt \\
\propto &  \int_{-\infty}^{+\infty} \dfrac{e^{-i x t} }{\left[ (t-\dfrac{i}{(1-\rho)\sa\sb})(t+\dfrac{i}{(1+\rho)\sa\sb}))\right]^{N/2}} dt, \nonumber
\end{align}
and perform the change of variable $t \to t+\dfrac{i\rho}{(1-\rho^2)\sa\sb}$
to obtain
\begin{equation}
  f_X(x) \propto e^{\dfrac{\rho x}{(1-\rho^2)\sa\sb}} \int_{-\infty}^{+\infty} \dfrac{e^{-i x t} }
{\left[ t^2+\dfrac{1}{(1-\rho^2)\sa\sb^2} \right]^{-N/2}} dt.
\end{equation}
Then making use of the Basset integral \citep[][Eq.10.32.11]{NIST2010} and reintroducing the normalization, we get: 
\begin{align}
\label{eq:pdfsum}
  f_X(x) =& \dfrac{ \abs{x}^{(N-1)/2}e^{\dfrac{\rho x}{(1-\rho^2)\sa\sb}} K_{(N-1)/2}\left( \dfrac{\abs{x}}{(1-\rho^2)\sa\sb}\right)} { 2^{(N-1)/2}\sqrt{\pi} \Gamma(N/2) \sqrt{1-\rho^2} (\sa\sb)^{(N+1)/2},  }
\end{align}
where $\Gamma$ refers to the gamma function and $K_\nu$ is the modified Bessel function of second kind and order $\nu = {(N-1)/2}$.
We can check a-posteriori that we recover indeed \refeq{prod1} for the $N=1$ case, which justifies \refeq{char1}.

We now have all in hands to consider a full-sky $A \times B$ cross-spectrum $\displaystyle{\hat \cl^{AB}=\dfrac{1}{2\ell+1} \sum_{l=-m}^{m}a_{lm}^A a_{lm}^{B\ast}}$ where for (isotropic) noise power ($N^A, N^B$) the covariance matrix reads
 \begin{equation}
\mathbf{V}=
\begin{pmatrix}
  \clth+N^A_\ell & \clth\\ 
  \clth &  \clth+N^B_\ell
\end{pmatrix}.
\end{equation}

Its \pdf for a given $\ell$ therefore reads:
\begin{equation}
  f_N(\hat \cl)= N f_X(N \hat \cl)
\end{equation}
where $N=2\ell+1$, and, in \refeq{pdfsum}: 
\begin{equation}
\begin{cases}
  \sa=\sqrt{\clth+N_\ell^A} \\
  \sb=\sqrt{\clth+N_\ell^B} \\
  \rho=\dfrac{\clth}{\sqrt{(\clth+N_\ell^A)(\clth+N_\ell^B)}}.
\end{cases}
\end{equation}
This formula is similar to the one given (but not derived) in
\citet[][Eq.19]{Percival:2006ss} for the TE distribution.

The characteristic function of the cross-spectrum estimator is:
\begin{equation}
  \label{eq:characfunc}
  \phi_{\hat C}(t)=\phi_X(\dfrac{t}{N}).
\end{equation}

%
\section{Auto-spectra and cross-spectra statistics comparison}\label{app:crossVSauto}
It is instructive to study the
respective merits of the auto and cross spectra approaches to estimate a single
field power-spectrum. We concentrate here on the full sky case where the auto and cross spectra estimators can be handled analytically.
The main conclusions hold essentially on a cut-sky too.

Let us first recall some properties of power-spectrum estimation
using auto-spectra. 
We consider the measurement of a Gaussian field (of power-spectrum \clth) over
the full sky with an instrument which has an isotropic noise (of power-spectrum
\Nc) uncorrelated to the signal.

According to the spectral theorem, the decomposition of the map onto
the (orthogonal) spherical harmonics basis
yields a set of \textit{independent} Gaussian random variables: \alm's.
For a given multipole $\ell$, this variable $x\equiv \alm$ follows
a Gaussian distribution of null mean and variance
$\sigma_\ell^2=\clth+N_\ell$.
At a given $\ell$ , from a set of $N=2\ell+1$ measured
harmonic coefficients $\{x_i\}$ the Maximum Likelihood
estimator of \clth\ is the empirical variance:
\begin{equation}
\hat \cl = \dfrac{1}{N}\sum_{i=1}^\ell x_i^2 -\Nc
\end{equation}

This estimator's distribution can be computed analytically by noticing
that since $x_i$ follows $\mathcal{N}(.;0,\clth+\Nc)$ it can be written as 
\begin{equation}
\label{eq:scaledchi2}
  \hat \cl =\dfrac{\clth+\Nc}{N}y - \Nc
\end{equation}
where $y$ follows now a $\chi^2_N$ distribution.
Then from the analytic $\chi^2$ \pdf and standard probability
transformation rules one can obtain analytically the full $\hat \cl$
\pdf (which is a $\Gamma$ one).
\footnote{We emphasize we are dealing for the moment with the estimator distribution, not
the posterior or likelihood one.}

We do not need any elaborate expression to compute the first order
moments. Instead we use the scaling property of the cumulants (which are equivalent to central moments up to the third order) which 
states that if $X$ has some cumulants $\kappa_i(X)$, the cumulants
of a linear transformation $Y=aX+b$ are $\kappa_i(Y)=a^i \kappa_i(X)
+b\delta_{i}^1$. 
Since the first cumulants of a $\chi^2_N$ distribution are
$\vec\kappa=N(1,2,8...)$, \refeq{scaledchi2} gives immediately:
\begin{align}\label{eq:auto-cumulants}
\kappa_1(\hat \cl)&=\E{\hat\cl}=\dfrac{\clth+\Nc}{N}N-\Nc=\clth  \\
\label{eq:varauto}
\kappa_2(\hat \cl)&=\textrm{Var}[{\hat\cl}]=\left(\dfrac{\clth+\Nc}{N}\right)^2 2N=\dfrac{2(\clth+\Nc)^2}{N}  \\
\kappa_3(\hat \cl)&=\E{(\hat \cl-\clth)^3}=\left(\dfrac{\clth+\Nc}{N}\right)^3 8N=\dfrac{8(\clth+\Nc)^3}{N^2} ,
\end{align}

Any error on the noise level estimation \Nc  bias accordingly the estimator
($\kappa_1$) and we can recognize the cosmic variance expression
($\kappa_2$).


The clear advantage of using cross-spectra has already been
emphasized: the estimator is unbiased whatever
knowledge we have of the noise spectra, \cf \refeq{meancross} .

However in order to investigate its discriminating power, we also need to
consider its variance,  and to a lesser extent its higher order
moments. What do we loose statistically using cross-spectra over using
an auto-spectrum assuming a perfect knowledge of the noise?
The answer depends on the relative levels of the signal (\clth) and
noise spectra  $(\Na,\Nb)$ and on the $\ell$ range under consideration.
To get some further insight, we consider the EE field from the \planck\ 2015 best-fit \citep{planck2015-XIII} and
variety of realistic maps noise-levels.
We focused on three particular public datasets: \wmap\ (V band), \lfi\ (70GHz) and two \hfi\ channels (100GHz and 143GHz).

We then consider the variance of the cross-spectrum estimator \refeq{varcross}
and compare it to the one obtained on the auto-spectrum of an
optimally inverse-variance combined map, \ie with noise
\begin{equation}
\dfrac{1}{\Nc}=\dfrac{1}{\Na}+\dfrac{1}{\Nb}
\end{equation}

\begin{figure}
  \centering
  \includegraphics[width=\linewidth]{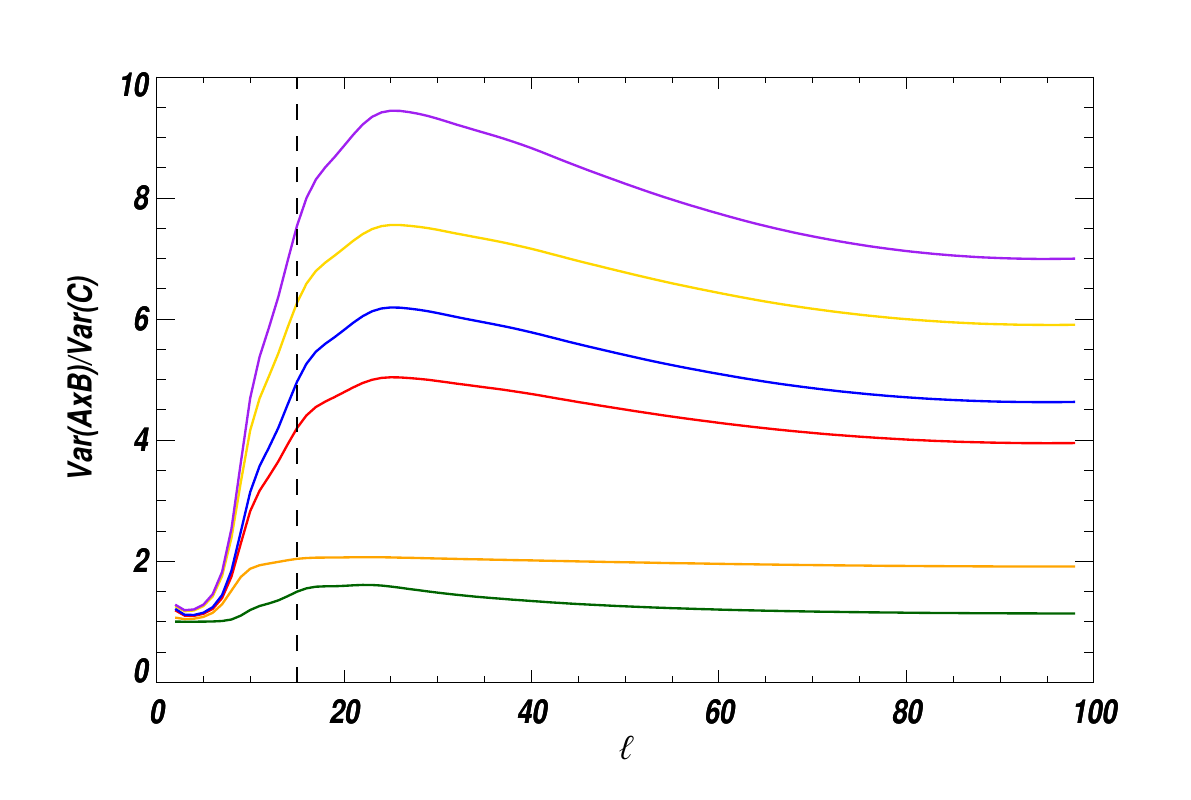}
  \caption{\label{fig:compvar} Ratio of the variance of the $A-B$ coss spectrum estimator
  to the one from the auto-spectrum of the optimally combined map. The
colors indicate different noise combinations according to the following
scheme: \colorcode. The dashed line recalls that most of the
interesting information in EE about reinization is contained below $\ell \lesssim
15$.}
\end{figure}

We show the ratio of these quantities  for the different noise-pair combinations
on Fig. \ref{fig:compvar}.

The variance increase is important in the case of two very
different noise levels (for instance \wmap$\times$\hfi) and moderate
when they are similar (\wmap$\times$\lfi, \planck-100$\times$\planck-143).
This can be understood from the
variance formulas Eqs. (\ref{eq:varauto}) and (\ref{eq:varcross}) where when $\Nb
\gg \Na, \Nc\simeq \Na$ :
\begin{align}
\label{eq:varlarge1}
  \mathrm{Var}(\clAB) \simeq & \dfrac{1}{N} \left(2(\clth)^2+\Nb\clth +\Na\Nb\right) \\
\label{eq:varlarge2}
  \mathrm{Var}(\hat \cl^C) \simeq & \dfrac{1}{N} \left(2(\clth)^2+4\Na\clth +2(\Na)^2\right).
\end{align}
Beyond the first term (the cosmic variance) which is dominant for
low-$\ell$'s, the cross-spectrum picks up 
the noisiest of the two measurement while auto-spectra uses
essentially the best one.

On the other side, when both measurements have similar noise levels,
$\Na\simeq\Nb,\Nc\simeq \dfrac{\Na}{2}$, the variances become similar:
\begin{align}
  \mathrm{Var}(\clAB) \simeq & \dfrac{1}{N} \left(2(\clth)^2+2\Na\clth +(\Na)^2\right) \\
  \mathrm{Var}(\hat \cl^C) \simeq & \dfrac{1}{N} \left(2(\clth)^2+2\Na\clth +\dfrac{(\Na)^2}{2}\right).
\end{align}

\begin{figure}
  \centering
  \includegraphics[width=\linewidth]{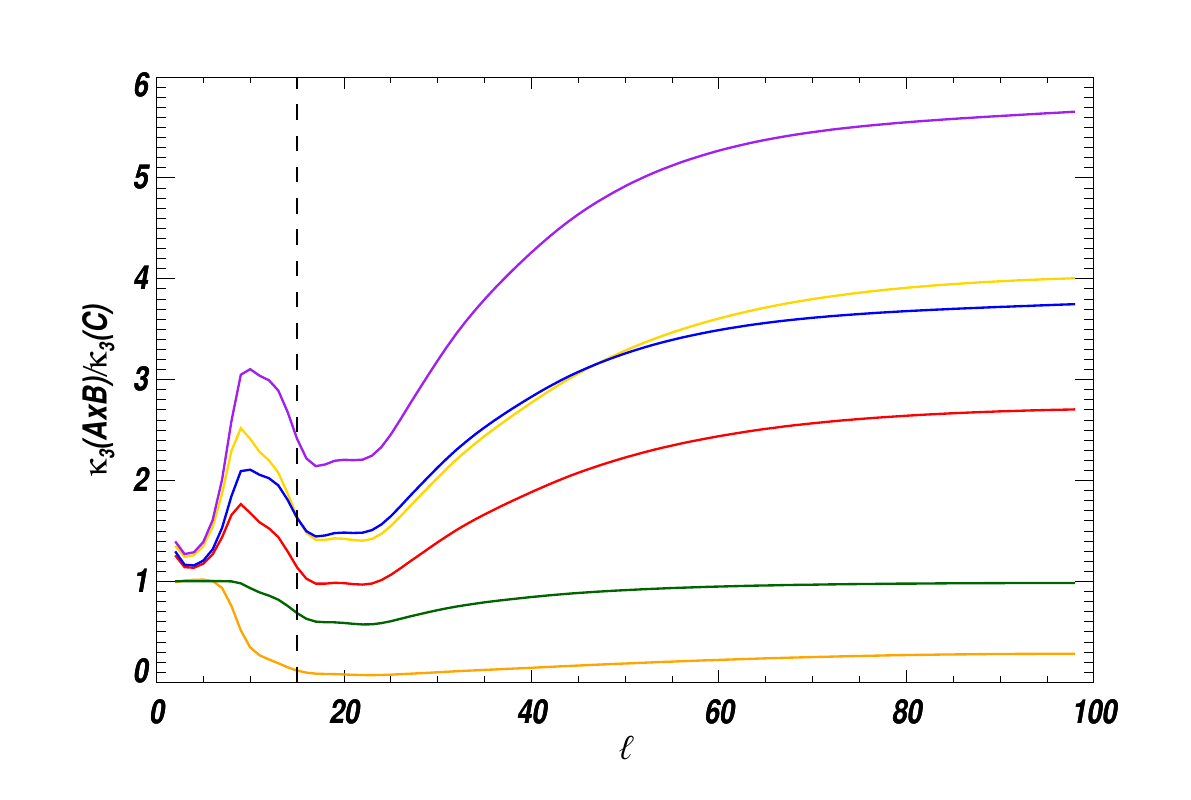}
  \caption{\label{fig:compskew} Ratio of the third order central
    moment ($\kappa_3$) for the $A\times B$ coss spectrum estimator
  to the one from the auto-spectrum of the optimally combined map. The
colors indicate different noise combinations according to the following
scheme: \colorcode. The dashed line recalls that most of the
intresting information in EE about reoinization is contained below $\ell \lesssim
15$.}
\end{figure}

Wether the linear term on \clth dominates or not depends on the signal, the noise levels and the $\ell$ range. 
As a rule-of-thumb, the comparision of \refeq{varlarge1} to
\refeq{varlarge2} suggests that the statistical
loss for a cross-combination is \q{reasonable} when the two 
noise levels are within a factor $\simeq 3$. The same kind of conclusion holds for the third central moment, but
one can get a smaller $\kappa_3$ value using cross-spectra for similar noise levels
(Fig. \ref{fig:compskew}).

\newpage
\section{\pdf parametrization}
\label{app:fit_distribs}

\begin{figure*}
	\centering
	\includegraphics[width=\linewidth]{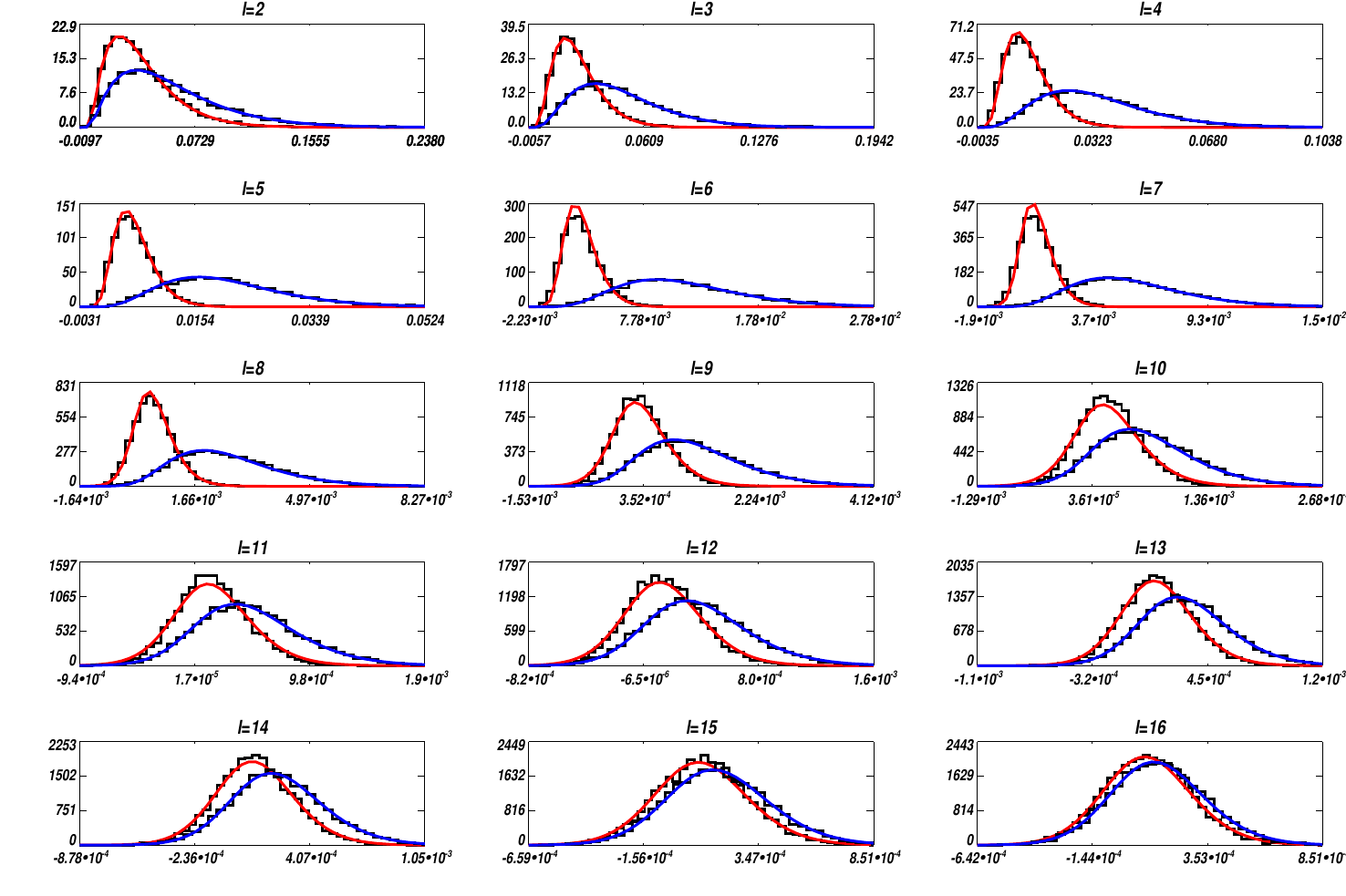}
	\caption{\label{fig:distrib5_anal} Normalized histograms (in black) of the 100$\times$143 PCL estimator in the $\ell \in [2,16]$ range are compared to our analytic full-sky based description, for model1 (blue) and model2 (red). The number of degree of freedom  $N(\ell)=(2\ell+1)\fskyc(\ell)$ is reduced according to the values obtained on model 2 only (Fig. \ref{fig:fsky_pc}).}
\end{figure*}

\begin{figure*}
  \centering
  \includegraphics[width=\linewidth]{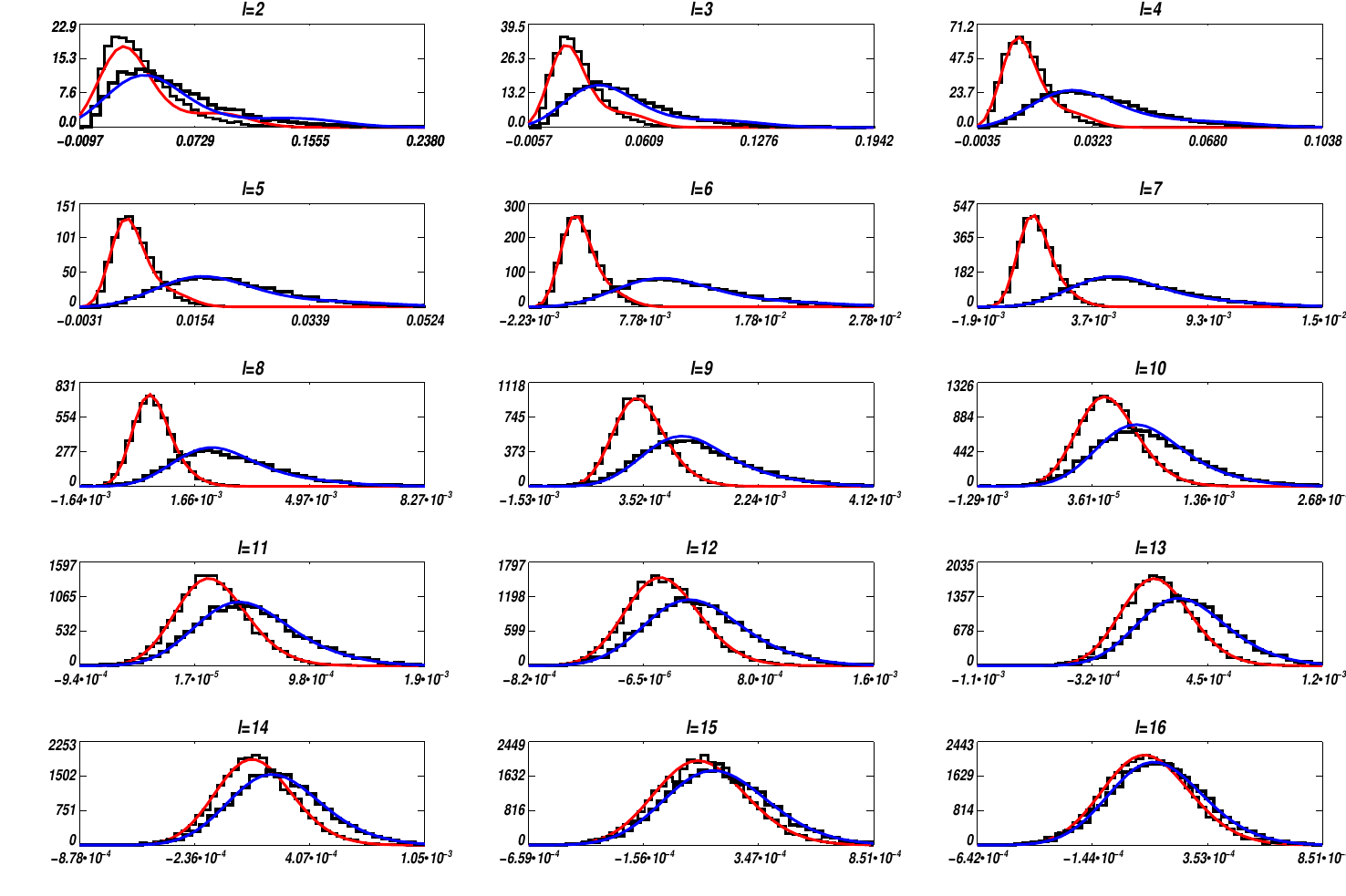}
\caption{\label{fig:distrib5_edge} Normalized histograms (in black) of the
  100$\times$143 PCL estimator in the $\ell \in [2,16]$ range are
  compared to our analytical parametrization based on the Edgeworth
  expansion for model1 (blue) and model2 (red). 
}
\end{figure*}

In this section we show the excellent agreement that is obtained when comparing the 
parametrization of the EE PCL estimator distribution defined with the full sky based approach and the Edgeworth expansion method described in \sect{subsec:result_analytic} and \sect{subsec:results_edge}, respectively. We consider the small mask with $f_{sky}=0.8$ and 
the \planck-100$\times$\planck-143 cross-spectrum simulations. Note that due to its low noise levels, this cross-spectrum is the most challenging to
describe. All other cross-spectra show an even better agreement. The results are summarized in \fig{fig:distrib5_anal} and \fig{fig:distrib5_edge}.

\label{sec:appendix}

\end{document}